 \documentclass[twocolumn,trackchanges,twocolappendix]{aastex631}

\usepackage{hyperref}
\usepackage{amsmath, amssymb}
\usepackage{float}

\begin{document}

\title{Spectral Analysis of Cyg X-3 using simultaneous \emph{AstroSat} and \emph{Insight}-HXMT Observations \footnote{Released on March, 1st, 2021}}

\author[0009-0004-9795-9820]{Suraj K. Chaurasia}
\affiliation{Department of Physics, Institute of Science, Banaras Hindu University, Varanasi-221005, India}

\author{Gitika Mall}
\affiliation{Center for Astronomy and Astrophysics, Center for Field Theory and Particle Physics and Department of Physics, Fudan University, Shanghai 200438, People's Republic of China}

\author{Ruchika Dhaka}
\affiliation{Department of Physics, IIT Kanpur, Kanpur, Uttar Pradesh 208016, India}

\author{Ranjeev Misra}
\affiliation{Inter-University Centre for Astronomy and Astrophysics (IUCAA), PB No.4, Ganeshkhind, Pune-411007, India}

\author{Amit Pathak}

\affiliation{Department of Physics, Institute of Science, Banaras Hindu University, Varanasi-221005, India}



\begin{abstract}

We present the results from the spectral analysis of Cygnus X-3 using simultaneous data from AstroSat and Insight-HXMT during its soft state. A pure reflection spectrum, including emission lines of iron, silica, and sulfur, provides a good fit to the spectra. Orbital phase-resolved analysis shows no significant spectral parameter variations, except for the normalization. Leveraging IXPE polarization results, we model the funnel-shaped geometry and estimate scattered flux and observed polarization for various funnel parameters and observer inclinations. We consider two scenarios: reflection from the funnel walls and scattering by gas within the funnel. Our results reconfirm previous findings, showing that reflection can produce a high polarization degree (PD) of 23$\%$, but not a low PD of 12$\%$. Conversely, scattering can produce a PD of 10-12$\%$, but not as high as 23$\%$ for a fixed observer inclination of $30^\circ$. Scattering results align with previous findings without absorption, but with absorption, PD drops significantly with increasing funnel opening angle. Thus, we can identify common funnel parameters that can produce the different observed PDs in the soft and hard states. The intrinsic luminosity of the source was estimated by comparing the results from a plane disk and the funnel model, to be  $\sim7$ $\times$ 10$^{40}$ erg/s for 12$\%$ PD (scattering) and $\sim5$ $\times$ 10$^{41}$ erg/s for 23$\%$ PD (reflection). However, for the reflection model, the luminosity may decrease to $\sim$ 10$^{40}$ erg/s when the 23$\%$ PD observed is taken as a lower limit.
\end{abstract}
\section{Introduction}
Cygnus X-3 is a Galactic high-mass X-ray binary system. It is one of the brightest, persistent, and most puzzling sources of X-ray emissions discovered back in 1966 (\citealt{1967ApJ...148L.119G}). Precise determination of the distance to Cygnus X-3 is challenging due to significant absorption in the Galactic plane, which obscures optical wavelengths. However, a recent study using the parallax method with Very Long Baseline Interferometry (VLBI) has reported the distance to the source to be 9.67 $\pm$ 0.5 kpc (\citealt{Reid_2023}). The mass donating companion of the system is a Wolf-Rayet (WR) star (\citealt{vanKerkwijk1992}). The inclination of Cyg X-3 is constrained to be around 30$^\circ$ as determined through photoionization simulations and analysis of orbital modulation of emission lines (\citealt{refId2}). \citealt{Antokhin_2022} reported the inclination of 29.5$^\circ$ $\pm$1.2$^\circ$ by analyzing the X-ray and IR light curves. The X-ray emissions from the source exhibit a pronounced 4.8-hour cycle of intensity variation (\citealt{1986ApJ...309..700M}), which is in direct correspondence with the orbital period of the system. Longer modulations in both the X-ray and radio lightcurves are linked to the precessional movement of the accretion disk (\citealt{2001ApJ...553..766M}). The nature of the compact object is not certain, but it is conjectured to be a black hole based on its spectral similarities to other X-ray binary systems associated with black holes such as GRS 1915+105 and XTEJ1550-654 (\citealt{10.1111/j.1365-2966.2008.13479.x}). By assessing the system's X-ray and radio flux in conjunction with spectral hardness, five distinct X-ray spectral states have been observed: Hypersoft state (HPS), Flaring soft X-ray state (FSXR), Flaring intermediate state (FIM), Flaring hard X-ray state and the quiescent state (\citealt{10.1111/j.1365-2966.2010.16722.x}). The Hardness-Intensity diagram (HID) serves as a crucial tool for comprehending the characteristics of transient black hole systems. In the typical transient outbursting cycle, the HID exhibits a Q-type shape (\citealt{2004MNRAS.355.1105F}). However, in the case of Cygnus X-3, there is an absence of hysteresis in the HID. Instead, the intensity of Cygnus X-3 simply increases without showcasing the typical behavior of spectrum softening (\citealt{10.1111/j.1365-2966.2008.14036.x}). 
\\
Recent study reveals high polarization of the source (\citealt{Veledina2024}, \citealt{veledina2024ultrasoftstatemicroquasarcygnus}) in both hard and soft states. The study measured X-ray polarization from Cygnus X-3, finding high linear polarization of 20.6$\%$ orthogonal to the direction of radio ejections during the hard state. The PD remains consistent across the 3.5-6.0 keV range but decreases in 6.0-8.0 keV, where the Fe K$_\alpha$ emission lines dominate. The geometry of the system implies the presence of an optically thick medium above the orbital plane, shaped like a funnel with its half-opening angle less than 15$^\circ$. The intrinsic X-ray luminosity exceeds the Eddington limit for a neutron star or a black hole, depending on the funnel's opening angle, placing Cygnus X-3 in the ULX class.
\\
In the ultrasoft state, \cite{veledina2024ultrasoftstatemicroquasarcygnus} observed a high polarization degree (PD) of 11.9 ± 0.5\% with a polarization angle (PA) of 94$^\circ$ ± 1$^\circ$, indicating that the central source is shrouded by a thick medium. This finding is consistent with the characteristics observed in the hard and intermediate states, suggesting a persistent envelope around the source across various states.
\\

In this paper, we perform spectral analysis of Cygnus X-3 using X-ray observations acquired from \emph{AstroSat} and \emph{Insight}-HXMT observatories, covering a broad energy spectrum of 1.0-20.0 keV. We performed the spectral analysis of the source throughout its entire orbital cycle and further examined specific orbital phases, including the declining phase, minimum phase, and rising phase. Our analysis focused on estimating the intrinsic luminosity and the polarization of scattered radiation, utilizing a funnel-shaped geometry under two distinct scenarios. In the first scenario, we examined scattering within the funnel's volume, which includes two cases: one where we considered absorption after scattering and another where absorption after scattering was ignored. In the second scenario, we considered reflection from the inner walls of the funnel. For a fixed observer inclination, we determined the observed reflected flux and the polarization degree (PD) for different funnel parameters across all cases and estimated the observed PD.

\section{Observation and data reduction}

\begin{figure}
    \includegraphics[width=2.5in,angle=270,trim=0 0 0 0,clip]{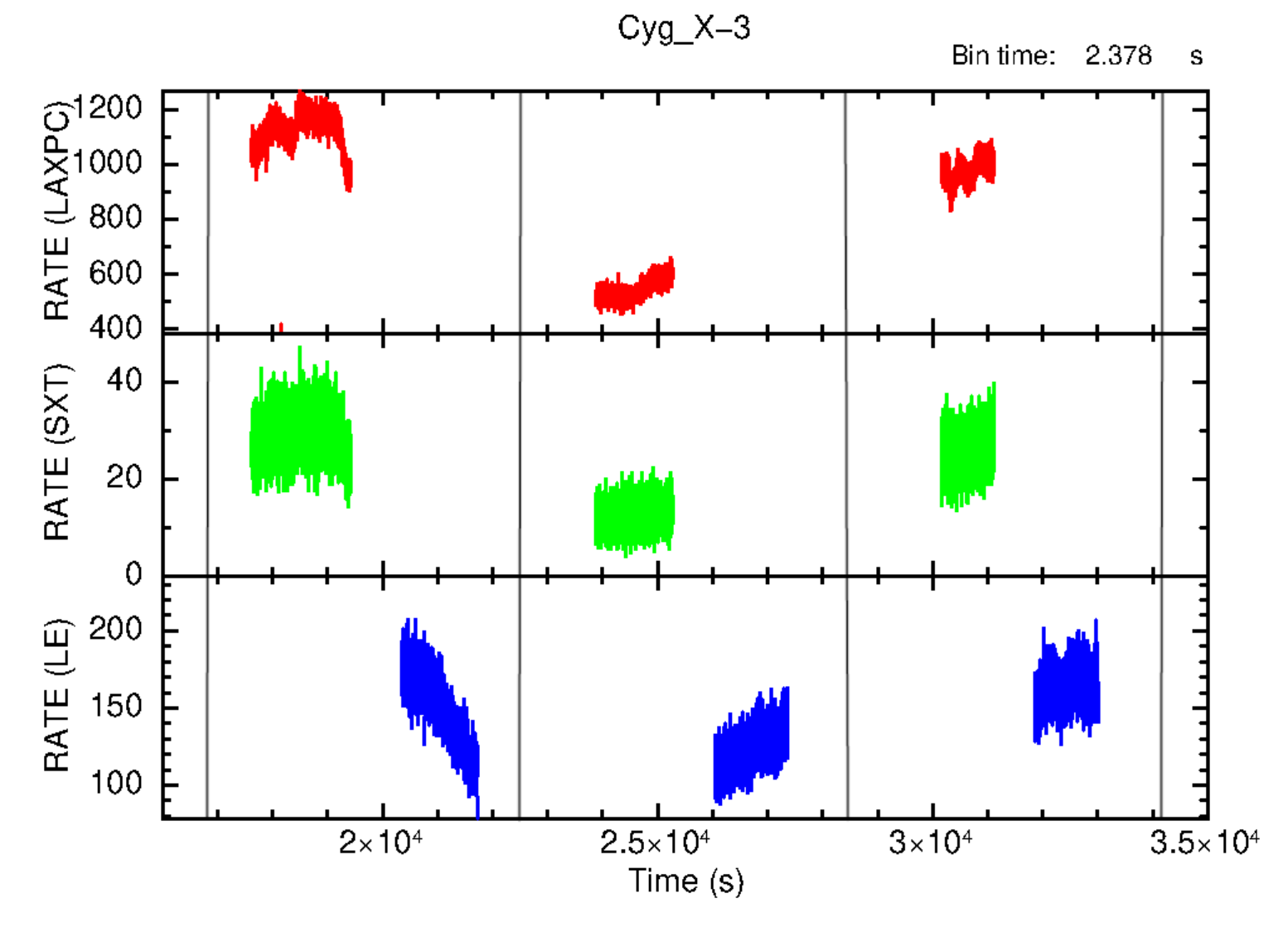}
    \caption{Simultaneous lightcurve of Cygnus X-3 observed by different instruments onboard \textit{AstroSat} (SXT and LAXPC) and \textit{Insight}-HXMT (LE). The lightcurve is plotted in counts per second, with Red, Green, and Blue markers showing LAXPC20, SXT, and LE lightcurves respectively. The grey vertical lines divide the complete orbit into three distinct orbital phases. }
    \label{fig:lightcurve}
\end{figure}

\textit{AstroSat} and \textit{Insight}-HXMT satellites observed Cygnus X-3 on August 6, 2018. The \textit{AstroSat} observation (ObsID: G08\_032T01\_9000002280) had an on source exposure time of 11.4 ks and \textit{Insight}-HXMT observation (ObsID: P0101298025) had an exposure of 20.6 ks. These observations allowed us to filter one orbit with simultaneous data from all the instruments onboard. We used one orbit data with 4.8 hour cycle to perform the phase resolved spectroscopy and divided the data into three orbital phases namely: declining phase, minimum phase, and the rising phase. The exposure times for each instrument during these phases are listed in Table \ref{tab:exposure}. The simultaneous lightcurve of a complete orbit from LAXPC, SXT, and LE is presented in Figure \ref{fig:lightcurve}. To facilitate the detection and comparison of the source's features observed by distinct instruments on different satellites, we implemented barycentric corrections to align the timestamps of individual photons with the reference frame of the solar barycenter. For spectral analysis covering the broad energy range of 1.0-20.0 keV, we utilized the LAXPC, SXT onboard \emph{AstroSat} and the LE, ME onboard \emph{Insight}-HXMT.

\setlength{\tabcolsep}{0.9pt}
\begin{table*}
\caption{Details of simultaneous observation of the Cygnus X-3 by \emph{AstroSat} and \textit{Insight}-HXMT. The observation IDs are listed alongside the exposure time for each orbital phase for different instruments.}
    \centering
    \begin{tabular}{c c c c c c}
    \hline \hline
    Observatory & Orbital Phase & Exposure (SXT) & Exposure (LAXPC) & Exposure (LE)& Exposure (ME)\\
    & &  (ks) &  (ks) &  (ks)&  (ks)\\
    \hline
    \textit{AstroSat}  (ObsID: G08\_032T01\_9000002280) & Declining & 1.8 & 1.8 & 1.4 & 1.0 \\
     \& & Minimum & 1.4 & 1.4 & 1.3 & 1.0 \\
    \textit{Insight}-HXMT (ObsID: P0101298025) & Rising & 1.0 & 1.0 & 1.1 & 0.8 \\
    \hline
    \end{tabular}
\label{tab:exposure}
    
\end{table*}

\subsection{AstroSat}

\begin{figure}
    \includegraphics[width=3.3in,angle=0,trim=0 0 0 0,clip]{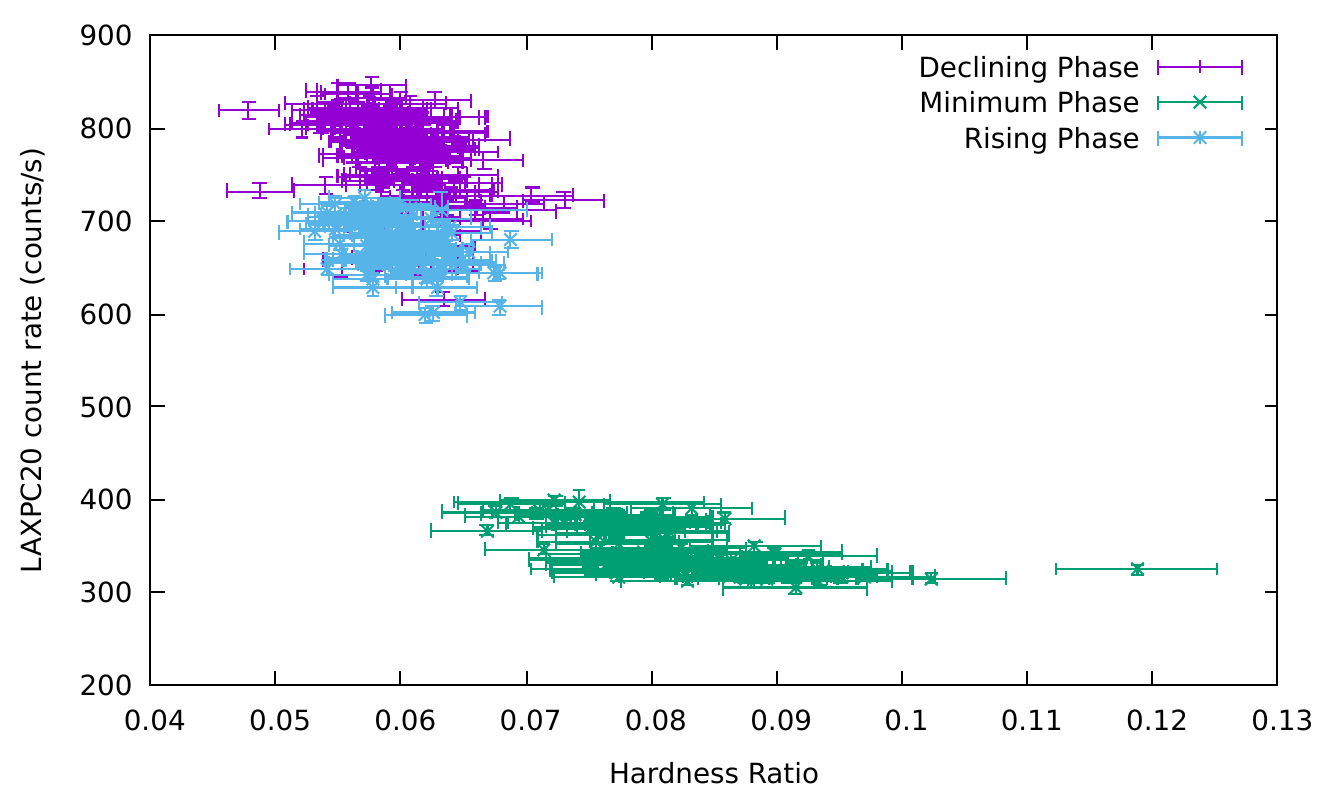}
    \caption{Hardness intensity diagram (HID) of Cygnus X-3 where 3-6 keV count rate is shown as a function of hardness ratio (ratio of count rate between 10-15 keV and 3-6 keV) using LAXPC20 lightcurve.}
    \label{fig:HID}
\end{figure}

 \textit{AstroSat} is a multi-wavelength observatory launched for astronomical studies of various celestial objects having four payloads: Soft X-ray Telescope (SXT), Ultra-Violet Imaging Telescope (UVIT), Large Area X-ray Proportional Counter (LAXPC) and Cadmium Zinc Telluride Imager (CZTI) (\citealt{2006AdSpR..38.2989A}). \\
 The Soft X-ray Telescope (SXT) is an imaging telescope working in the soft X-ray band of 0.3-8 keV energy range (\citealt{Singh2017}, \citealt{2016SPIE.9905E..1ES}). The SXT operates in two modes, Photon Counting (PC) mode and Fast Windowed Photon Counting (FW) mode.
We processed the Level-1 data using the SXT pipeline software (version: AS1SXTLevel2-1.4b), cleaned the Level-2 event files from different orbits, and merged them with the SXT event merger tool. The source spectrum was extracted from a circular region with a radius of 10 arcminutes, centered on the source coordinates. For the background spectrum, we used the blank sky SXT spectrum "SkyBkg\_comb\_EL3p5\_Cl\_Rd16p0\_v01.pha" and the “sxt\_pc\_mat\_g0to12.rmf” file as the redistribution matrix file (RMF). We generated off-axis auxiliary response files (ARF) using the \texttt{sxtARFModule} tool, providing the on-axis ARF "sxt\_pc\_excl00\_v04\_20190608\_mod\_16oct21.arf" as input. The individual energy spectra from all observations were re-binned using \texttt{ftgrouppha} using the Kaastra \& Bleeker optimal binning algorithm (\citealt{refId1}).
In modeling the spectra, we modified the gain of the response file using the \texttt{gain fit} command in XSPEC, fixing the slope at unity while leaving the offset as a free parameter. For our analysis, we utilized the SXT spectrum in the 1.0-7.0 keV energy range.\\
  The Large Area X-ray Proportional Counter (LAXPC) consists of three identical proportional counters known as LAXPC10, LAXPC20, and LAXPC30 operating in the wide energy range of 3-80 keV, providing a time resolution of 10 $\mu$s and deadtime of about 42 $\mu$s. (\citealt{Yadav_2016}, \citealt{Antia_2017}, \citealt{2017JApA...38...27A}). We used LAXPC software (\texttt{LaxpcSoft}; version as of 2022 August 15) to process the Level-1 Event Analysis (EA) mode data. The data reduction and the extraction of science products were carried out using standard tools available in \texttt{LaxpcSoft}. The Hardness-Intensity Diagram (HID) was obtained using LAXPC20 in the 3.0-6.0 keV and 10-15 keV energy bands, as shown in Figure \ref{fig:HID}. We used only the LAXPC20 detector in this study as LAXPC30 was switched off in 2018 March due to the gas leakage and LAXPC10 was operating at low gain (\citealt{Antia_2017}). We modeled the LAXPC spectrum of Cygnus X-3 in the 5-20 keV energy band, the high energy range i.e. $>$ 20.0 keV was ignored because of the low signal-to-noise ratio. \\

\subsection{Insight-HXMT}
\textit{Insight}-HXMT is a Chinese X-ray telescope, equipped with low-energy (LE), medium-energy (ME), and high-energy (HE) detectors that cover the energy range of 1-250 keV (\citealt{2020SCPMA..6349505C}, \citealt{2020SCPMA..6349503L}, \citealt{2020SCPMA..6349502Z}). We extracted the light curves and spectra following the official user guide\footnote{\url{http://hxmtweb.ihep.ac.cn/SoftDoc.jhtml }} using version 2.06 of the HXMTDAS software\footnote{\url{http://hxmtweb.ihep.ac.cn/software.jhtml }}. Background estimation was performed using the scripts hebkgmap, mebkgmap, and lebkgmap (\citealt{2020JHEAp..27...44G}, \citealt{liao2020background}). Good time intervals were screened based on the recommended criteria: elevation angle $>$ 10 degrees, geomagnetic cut-off rigidity $>$ 8 GeV, pointing offset angle $<$ 0.1 degrees, and at least 300 seconds away from the South Atlantic Anomaly (SAA).

For the analysis of Cygnus X-3, we modeled the spectrum obtained from the ME and LE instruments in the 10-20 keV and 2-10 keV energy bands, respectively. Due to the significant drop in flux above 20 keV, the HE spectra were excluded from the spectral study. To ensure optimal analysis of the X-ray data, we employed the Kaastra \& Bleeker optimal binning algorithm to group the LE and ME spectra using the tool \texttt{ftgrouppha} similar to SXT.

\section{Broadband X-ray Spectral Analysis}

\begin{figure*}
    \centering
    \includegraphics[width=2.3in ,angle=270,trim=0 0 0 0,clip]{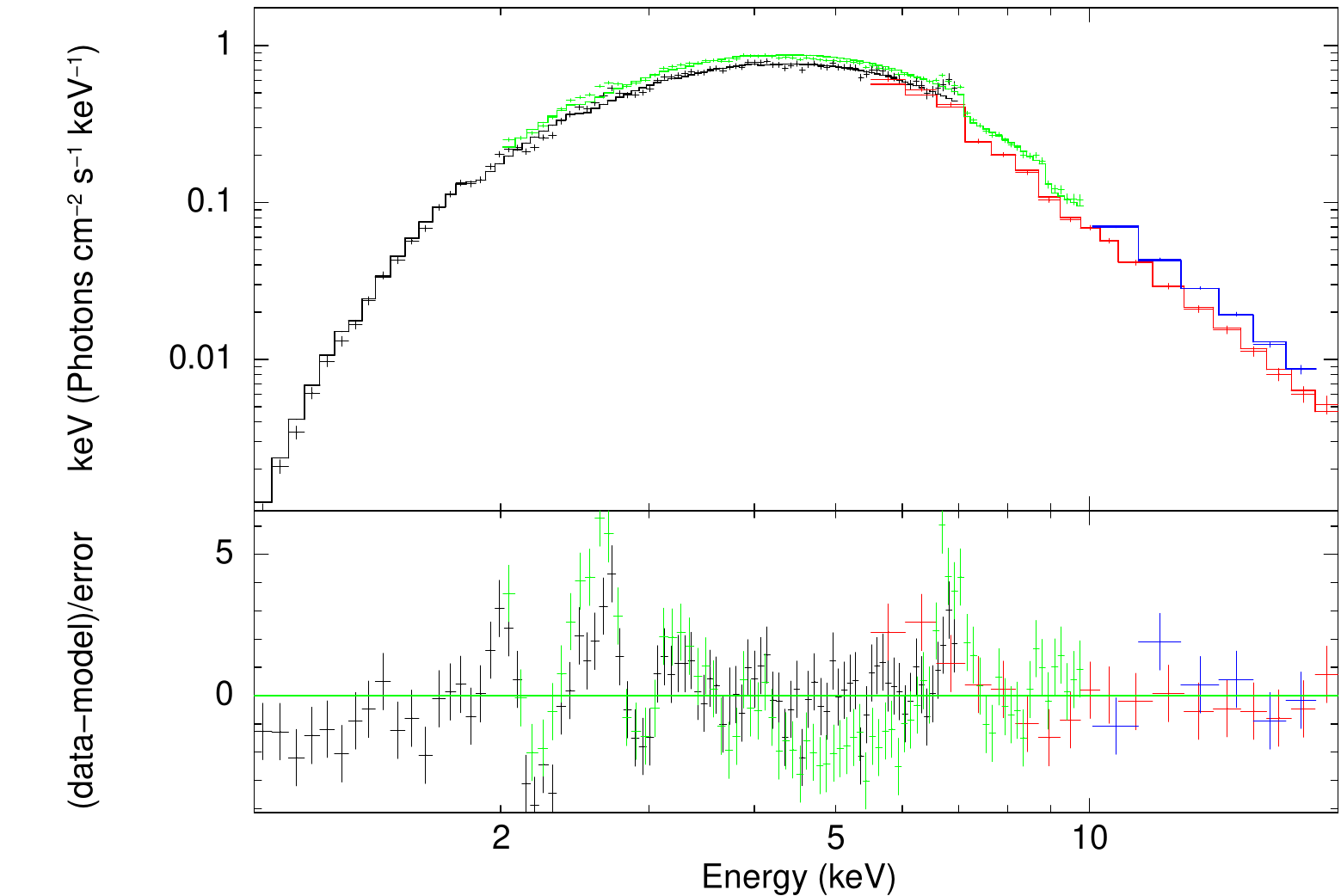}
    \caption{Fitted joint spectra along with model components and residuals for full orbit in 1-20 keV energy band. Spectra were fitted using the model \texttt{{tbabs}*{reflect}*{thcomp}*{diskbb}}. The black, red, green, and blue markers represent SXT (1.0-7.0 keV), LAXPC20 (5.0-20.0 keV), LE (2.0-10.0 keV), and ME (10.0-20.0 keV) data, respectively.}
    \label{fig:fig3}
\end{figure*}

\begin{figure*}
    \centering
    \includegraphics[width=2.3in,angle=270,trim=0 0 0 0,clip]{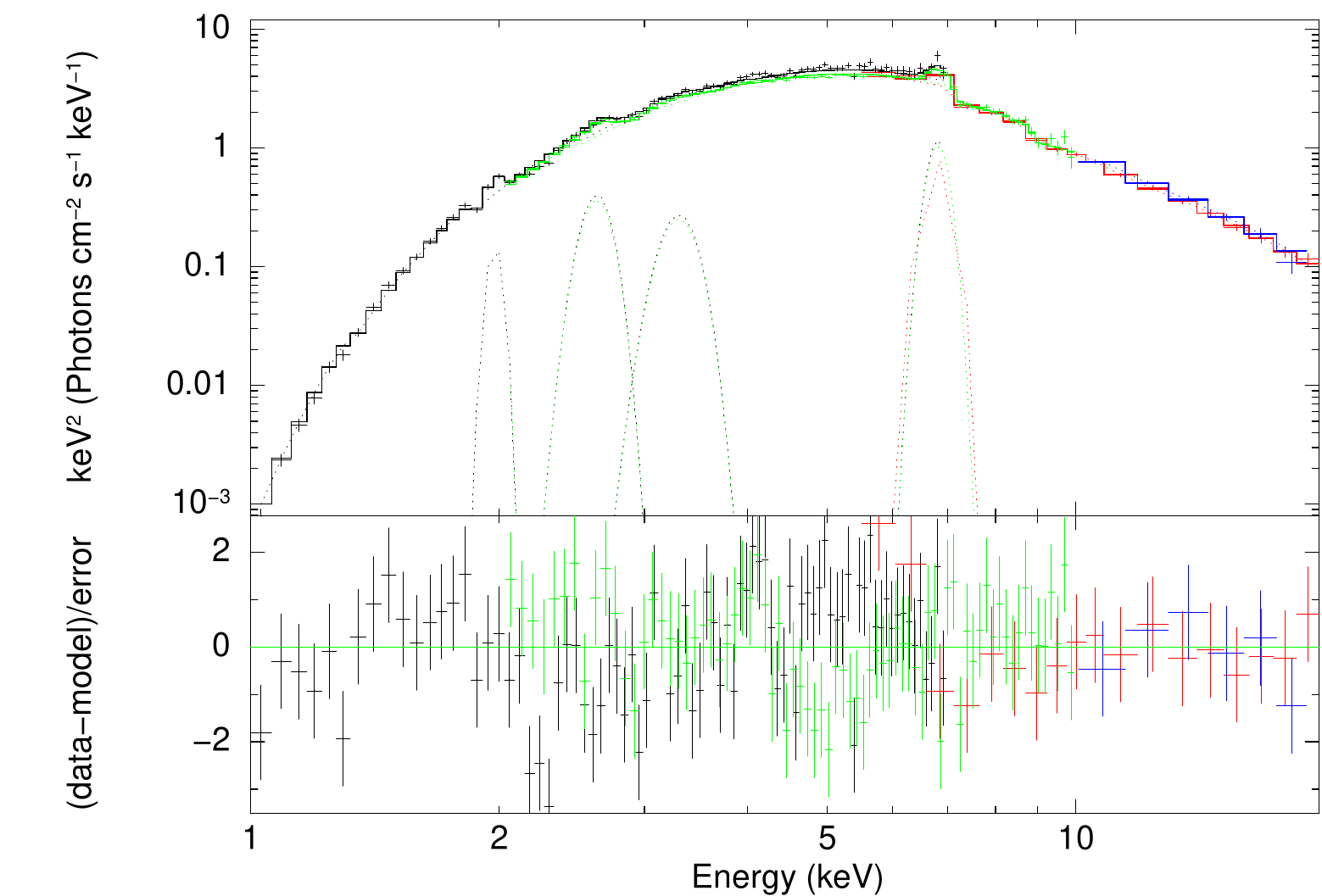}\includegraphics[width=2.3in,angle=270,trim=0 0 0 0,clip]{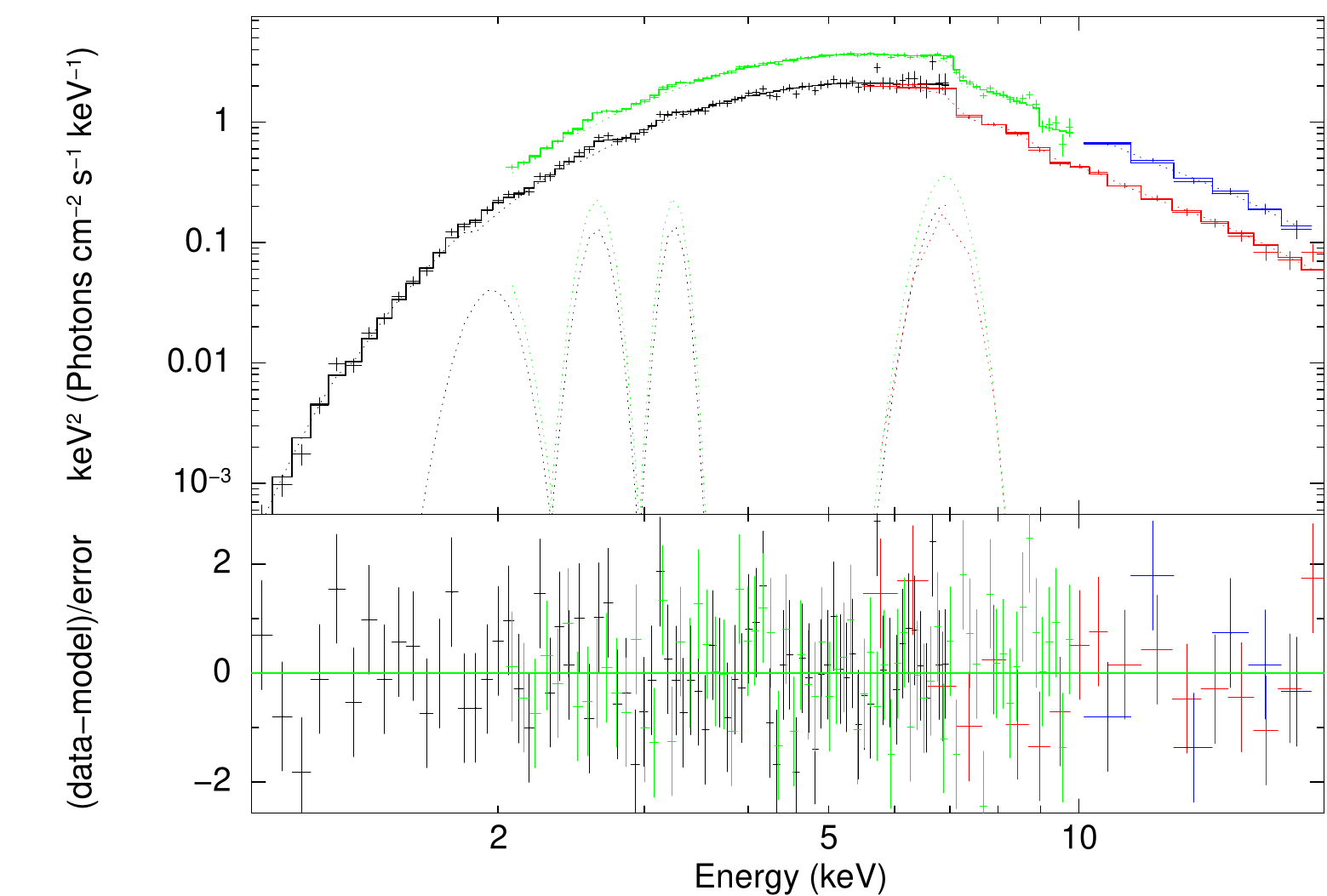} \includegraphics[width=2.3in,angle=270,trim=0 0 0 0,clip]{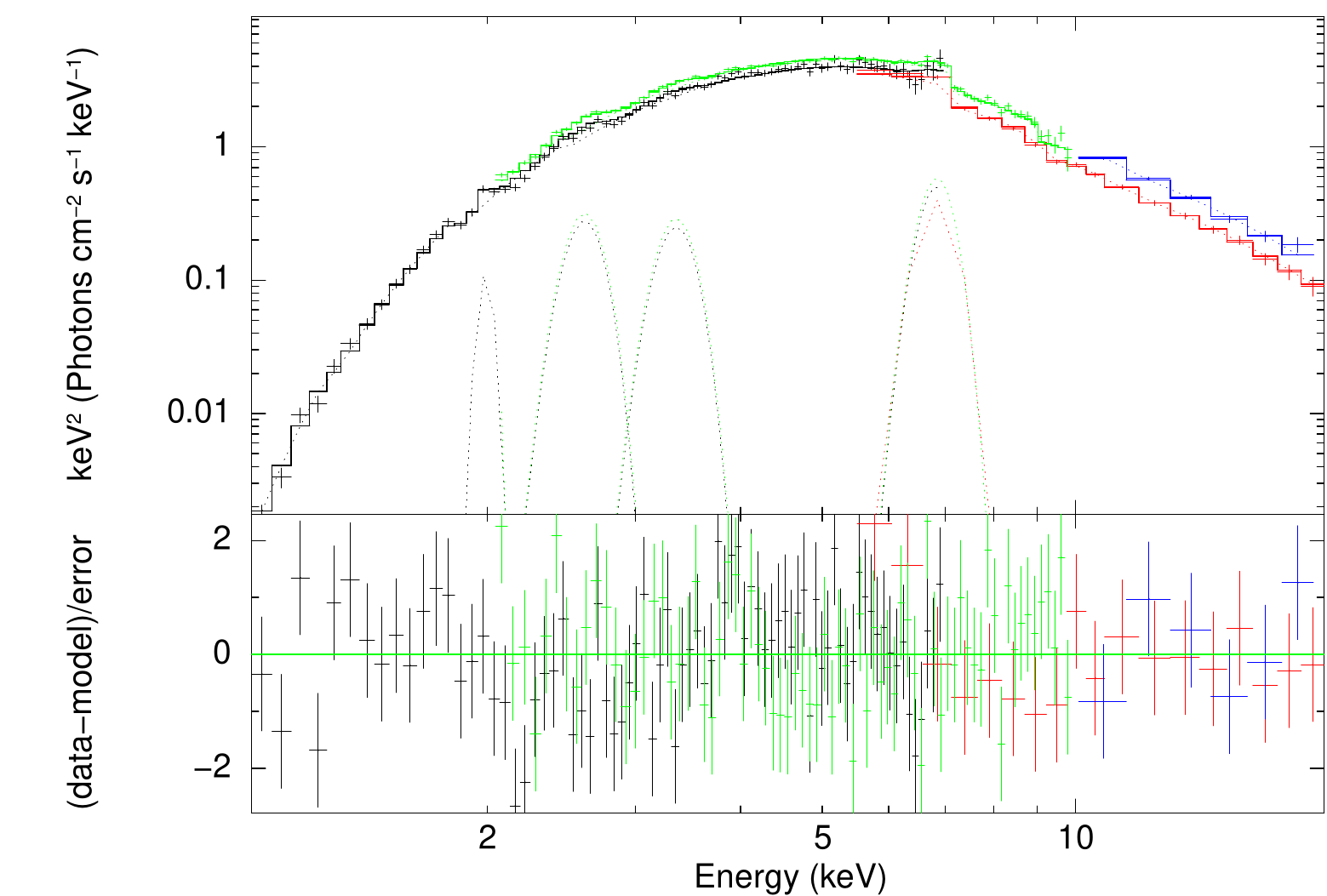}\includegraphics[width=2.3in,angle=270,trim=0 0 0 0,clip]{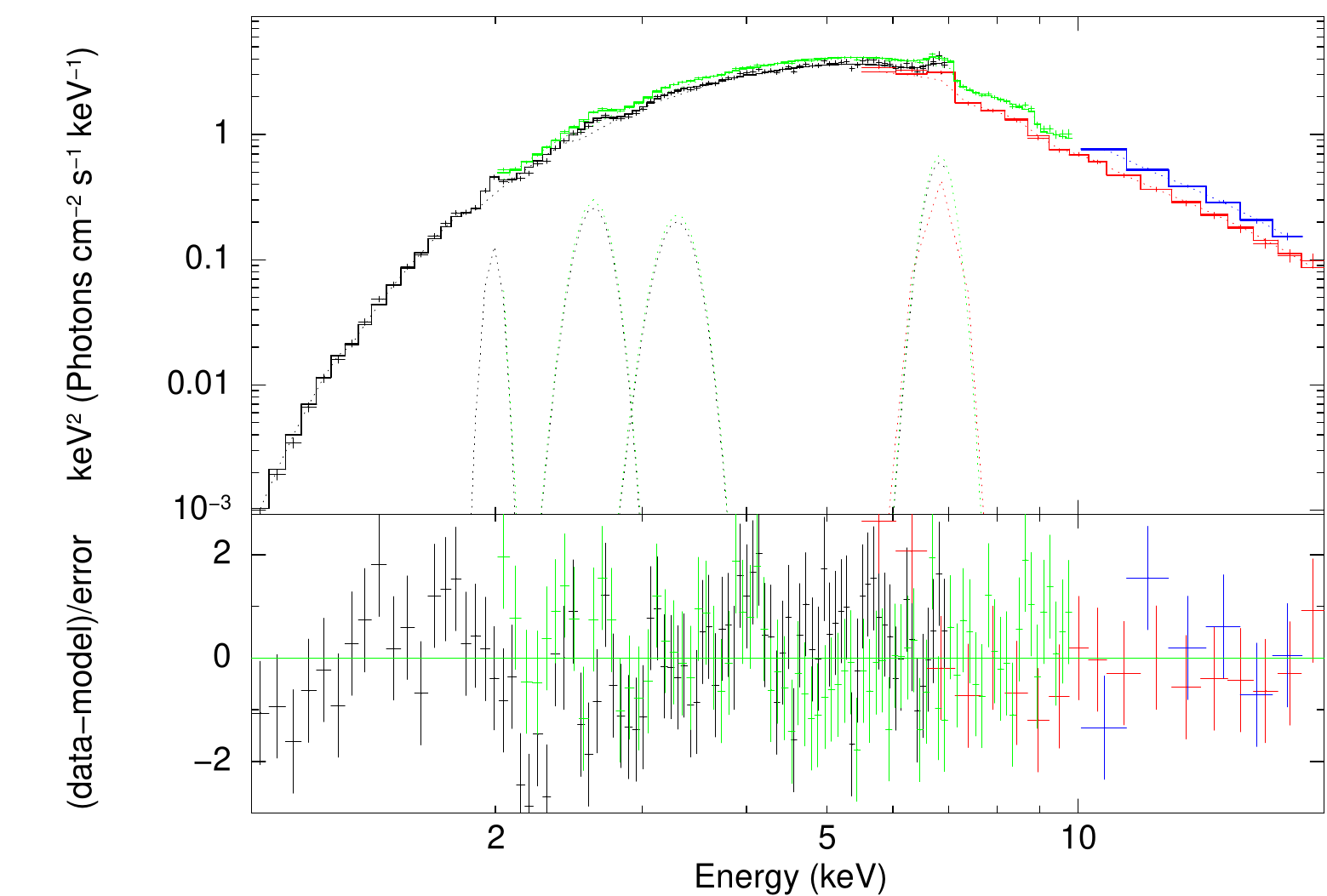}
    \caption{Top left and right panels show the data to best-fit model ratios for declining and minimum orbital phases respectively. The bottom left and right panels show the data to best-fit model ratios for the rising phase and for the complete orbit respectively using the model \texttt{tbabs*(reflect*edge*thcomp*diskbb + gaussian + gaussian + gaussian + gaussian)}. The black, red, green, and blue markers represent SXT (1.0-7.0 keV), LAXPC20 (5.0-20.0 keV), LE (2.0-10.0 keV), and ME (10.0-20.0 keV) data, respectively.}
    \label{fig:fig4}
\end{figure*}

In our study, we conducted the analysis of broadband X-ray spectra of Cygnus X-3 during its soft state. The study focused on the full orbit observation within the 1.0-20.0 keV energy band, employing the X-ray spectral fitting software XSPEC v12.13.0c (\citealt{1996ASPC..101...17A}). Inspired by the observed polarization characteristics, we explored a spectral model where the emission is primarily driven by reflection. As mentioned by \citealt{veledina2024ultrasoftstatemicroquasarcygnus}, the large polarization indicates that we are observing only the reflected spectrum (or the scattered one) and not the intrinsic source. We utilized the XSPEC model \texttt{reflect} (\citealt{1995MNRAS.273..837M}), parametrized by a solid angle $R = \frac{\Omega}{2\pi}$. In an unobscured geometry, the parameter $R$ corresponds to the true solid angle of the reflector as seen from the primary X-ray source. We employ the thcomp Comptonization model (\citealt{2020MNRAS.492.5234Z}). This model involves the upscattering of seed photons either by a hot corona in the inner region or on top of the cold accretion disk. The seed photons originate from an accretion disk characterized by the spectrum of a disk blackbody (\texttt{diskbb}; \citealt{1984PASJ...36..741M}). We accounted for interstellar medium (ISM) absorption by applying the model \texttt{tbabs} (\citealt{2000ApJ...542..914W}). To account for inter-instrument variations and calibration uncertainties, a multiplicative constant was introduced to the spectral model. This constant was fixed at 1.0 for the LAXPC spectrum, while for the SXT, LE, and ME spectra, it was left as a free parameter to vary. The disk inclination and the temperature of the corona (kT$_{\rm e}$) were fixed at 30$^{\circ}$ and at 50 keV respectively. Iron abundance was fixed at 0.3 (solar abundance). We fixed the parameter rel\_refl = -1 to obtain only the reflection component and exclude the contribution of incident X-ray emission in the resulting spectra. The complete model can be expressed as \texttt{{tbabs}*{reflect}*{thcomp}*{diskbb}} in XSPEC terminology which resulted in a reduced ${\chi^2}$ value of 3.0. The fitted model with data and residual is plotted in Figure \ref{fig:fig3}.
\\
Studies conducted by \cite{Kallman_2019} have exposed a rich line complex, featuring both absorption and emission characteristics in the energy range of 1-10 keV. The presence of a range of ionization species of iron in the Cygnus X-3 wind also contributes to a rich emission line spectrum. Also, the presence of iron line complex ($\sim$ 6.8 keV), S XVI lines ($\sim$ 2.6 keV, 3.2 keV) and a Si XIV line ($\sim$ 2.0 keV) can be seen from the residual plot of Figure \ref{fig:fig3}. \texttt{Reflect} does not include the Iron emission line and hence we have to add the line additionally. Subsequently, we increased the model complexity by incorporating four Gaussian lines. An additional edge component at $\sim 9.0$ keV is also required by the data. This resulted in a significant improvement in the fit, yielding a reduced ${\chi^2}$ value of 1.1. Our best-fitting model can be written in XSPEC terms as \texttt{{tbabs}*({reflect}*{edge}*{thcomp}*{diskbb} + {gaussian} + {gaussian} + {gaussian} + {gaussian})}.
The errors on the parameters are provided with a 90\% confidence level and their values are stated in Table \ref{tab:spectral_fit}. Additionally, the unfolded spectra along with residuals are depicted in Figure \ref{fig:fig4}.
\\
In order to address uncertainties in calibration, we incorporated additional systematic errors of 3\% for LAXPC and SXT spectra, and 2\% for LE spectra. No systematic was added to ME spectra.\\
We studied the variation in the spectral lines and other spectral parameters over the course of the orbital phase and found no significant variation in the parameters except normalization.
 \\[2pt]

\setlength{\tabcolsep}{10pt}
\begin{table*}

    \caption{Broadband spectral parameters for the full orbit data as well as the orbital phases using XSPEC model \texttt{tbabs*(reflect*edge*thcomp*diskbb + gaussian + gaussian + gaussian + gaussian)}. $N_{H}$ is the neutral hydrogen column density; rel\_refl is the reflection scaling factor; $\Gamma$ is asymptotic power-law index; $kT_{e}$ is electron temperature; $kT_{in}$ is the temperature at inner disk radius.}

    \centering
    \begin{tabular}{c c c c c}
    \hline \hline
    Parameter & Complete Orbit & Declining Phase & Minimum Phase & Rising Phase  \\
\hline 
$N_{H}$ ($10^{22}$ cm$^{-2}$) & $2.18_{-0.06}^{+0.07}$ & $2.22_{-0.07}^{+0.06}$ & $2.22_{-0.07}^{+0.15}$ & $2.23_{-0.09}^{+0.10}$\\
rel-refl (reflect) & -1$^f$ & -1$^f$ & -1$^f$ & -1$^f$\\
Edge Energy (keV)& $8.91_{-0.07}^{+0.07}$ & $8.87_{-0.12}^{+0.12}$ & $8.93_{-0.11}^{+0.13}$ & $8.99_{-0.15}^{+0.17}$\\
MaxTau (edge)& $0.32_{-0.06}^{+0.05}$ & $0.30_{-0.07}^{+0.07}$ & $0.35_{-0.12}^{+0.07}$ & $0.32_{-0.08}^{+0.08}$\\
$\Gamma$ (thcomp)& $5.75_{-0.24}^{+0.13}$ & $5.86_{-0.19}^{+0.25}$ & $5.49_{-0.54}^{+0.21}$ & $5.67_{-0.32}^{+0.20}$\\
Cov frac (thcomp)& $0.62_{-0.17}^{+0.13}$ & $0.69_{-0.16}^{+0.31}$ & $0.46_{-0.23}^{+0.13}$ & $0.54_{-0.18}^{+0.17}$\\
Thcomp $kT_e$ (keV) & 50$^f$ & 50$^f$& 50$^f$&50$^f$ \\ 
Diskbb $kT_{in}$ (keV) & $0.89_{-0.02}^{+0.03}$ & $0.87_{-0.03}^{+0.02}$ & $0.93_{-0.02}^{+0.03}$ & $0.88_{-0.02}^{+0.03}$\\
Diskbb Norm (10$^{3}$) & $25.6_{-2.8}^{+2.8}$ & $36.7_{-4.3}^{+7.5}$ & $12.0_{-1.8}^{+1.8}$ & $32.0_{-4.9}^{+4.9}$\\
S XVI line 1 energy (keV)& $3.29_{-0.05}^{+0.05}$ & $3.29_{-0.03}^{+0.06}$ & $3.25_{-0.08}^{+0.06}$ & $3.33_{-0.06}^{+0.06}$\\
S XVI line 1 sigma (keV)& $0.15_{-0.05}^{+0.06}$ & $0.16_{-0.05}^{+0.07}$ & $0.07_{-0.07}^{+0.06}$ & $0.16_{-0.05}^{+0.11}$\\
S XVI line 1 Norm (10$^{-3}$)& $8.1_{-2.6}^{+2.3}$ & $12.3_{-3.6}^{+4.2}$ & $3.0_{-1.3}^{+1.5}$ & $10.4_{-3.4}^{+4.4}$\\
Fe line energy (keV) & $6.80_{-0.04}^{+0.04}$ & $6.81_{-0.04}^{+0.04}$ & $6.74_{-0.13}^{+0.11}$ & $6.81_{-0.08}^{+0.08}$\\
Fe line sigma (keV) & $0.22_{-0.05}^{+0.06}$ & $0.19_{-0.04}^{+0.05}$ & $0.32_{-0.10}^{+0.14}$ & $0.28_{-0.10}^{+0.09}$\\
Fe line Norm (10$^{-3}$)& $6.7_{-1.2}^{+1.0}$ & $10.6_{-1.8}^{+1.9}$ & $4.2_{-1.2}^{+2.1}$ & $7.4_{-2.0}^{+2.6}$\\
Si line energy (keV)& $1.99_{-0.02}^{+0.02}$ & $1.97_{-0.02}^{+0.02}$ & $1.93_{-0.32}^{+0.07}$ & $1.99_{-0.04}^{+0.04}$\\
Si line sigma (keV)& $0.04_{-0.04}^{+0.03}$ & $0.03_{-0.03}^{+0.03}$ & $0.12_{-0.07}^{+0.24}$ & $0.03_{-0.03}^{+0.05}$\\
Si line Norm (10$^{-3}$)& $7.5_{-2.2}^{+2.4}$ & $9.9_{-3.5}^{+4.0}$ & $8.9_{-4.5}^{+5.3}$ & $7.0_{-3.7}^{+4.4}$\\
S XVI line 2 energy (keV)& $2.61_{-0.02}^{+0.02}$ & $2.61_{-0.02}^{+0.02}$ & $2.62_{-0.03}^{+0.03}$ & $2.59_{-0.03}^{+0.03}$\\
S XVI line 2 sigma (keV)& $0.10_{-0.02}^{+0.02}$ & $0.10_{-0.02}^{+0.02}$ & $0.09_{-0.04}^{+0.04}$ & $0.12_{-0.02}^{+0.03}$\\
S XVI line 2 Norm (10$^{-3}$)& $14.9_{-2.5}^{+2.3}$ & $22.7_{-3.6}^{+3.8}$ & $6.4_{-1.9}^{+4.3}$ & $19.0_{-3.7}^{+3.9}$\\
$\chi^2/d.o.f.$ & $182.74/169$   & $206.97/160$ & $149.47/152$ & $161.97/155$  \\ 
\hline

    \end{tabular}
    \tablecomments{$f$ denotes fixed parameters during the fitting.}    
    \label{tab:spectral_fit}
\end{table*}

\section{Funnel Model}

\begin{figure}
    \centering
    \includegraphics[width=2.6in,angle=0,trim=0 0 0 0,clip]{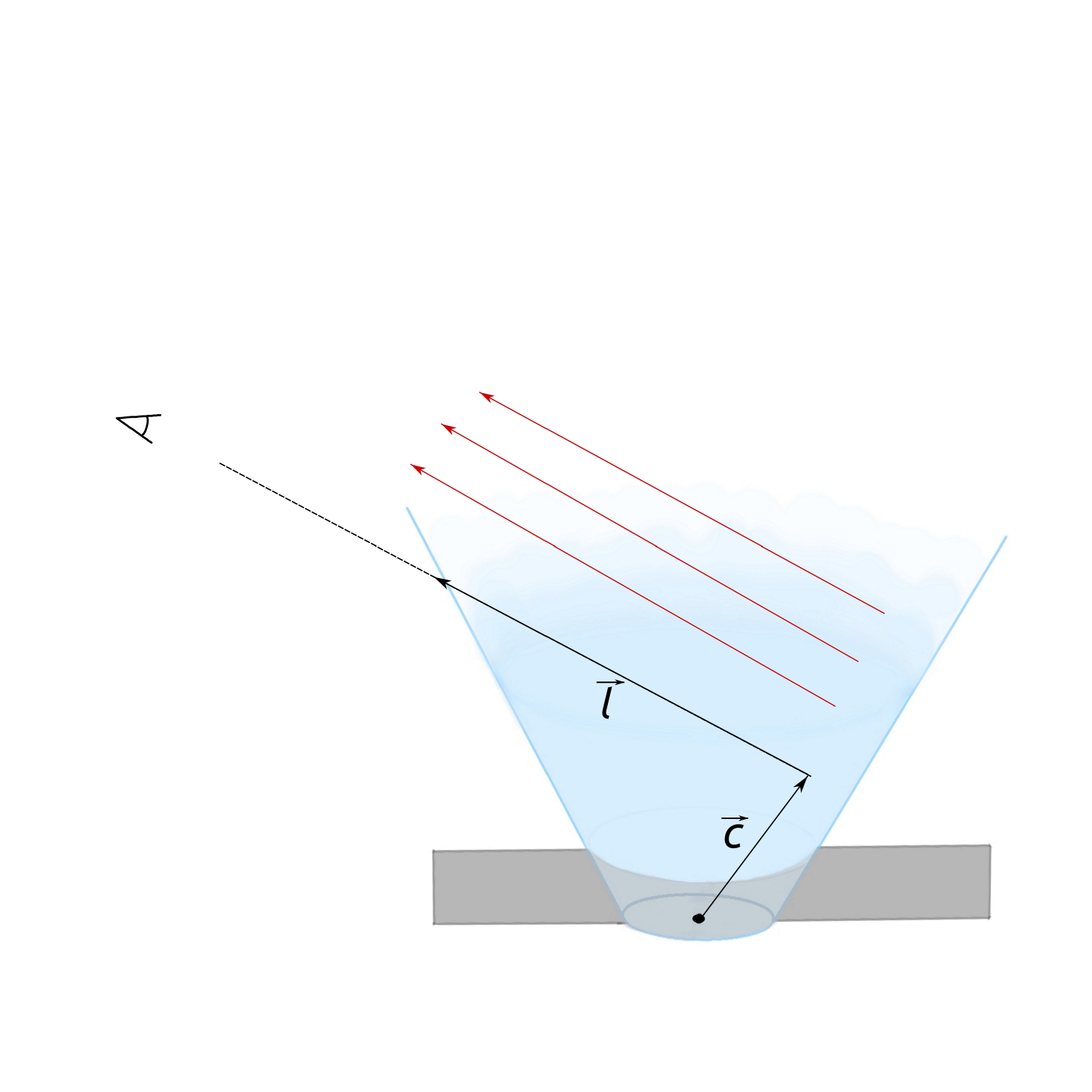}
    \caption{Above panel shows the funnel geometry with the central source marked with black sphere at the bottom and the observer at an inclination $i$ relative to the disk for the case of scattering taking place from the gas present inside the funnel volume. The gray shaded region shows the obstructed part of the funnel.}
    \label{fig:fig5}
\end{figure}

\begin{figure}
    \centering
    \includegraphics[width=3.4in,angle=0,trim=0 0 0 0,clip]{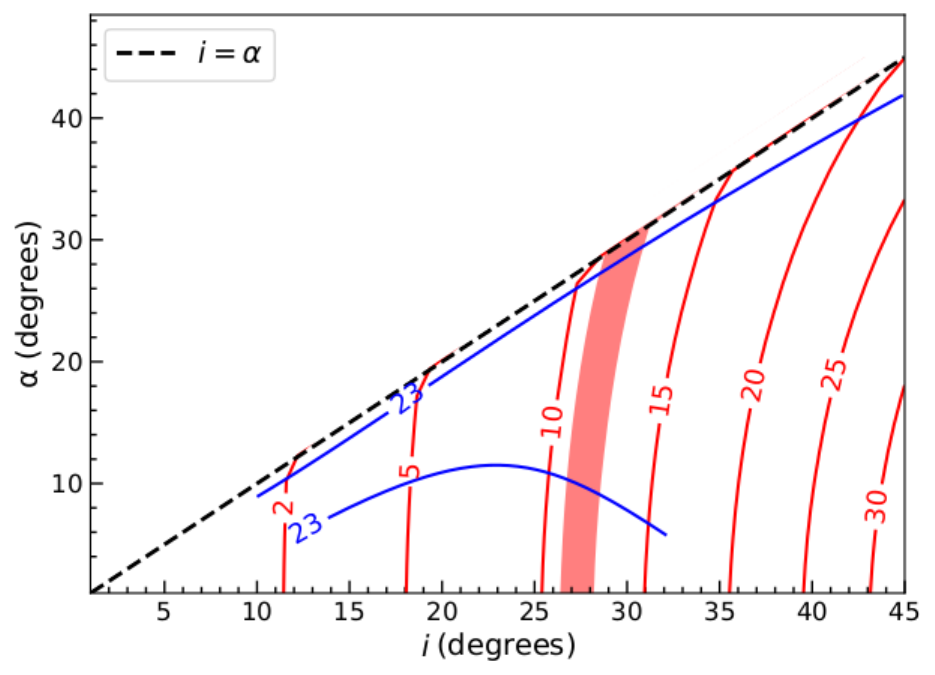}
    \includegraphics[width=3.4in,angle=0,trim=0 0 0 0,clip]{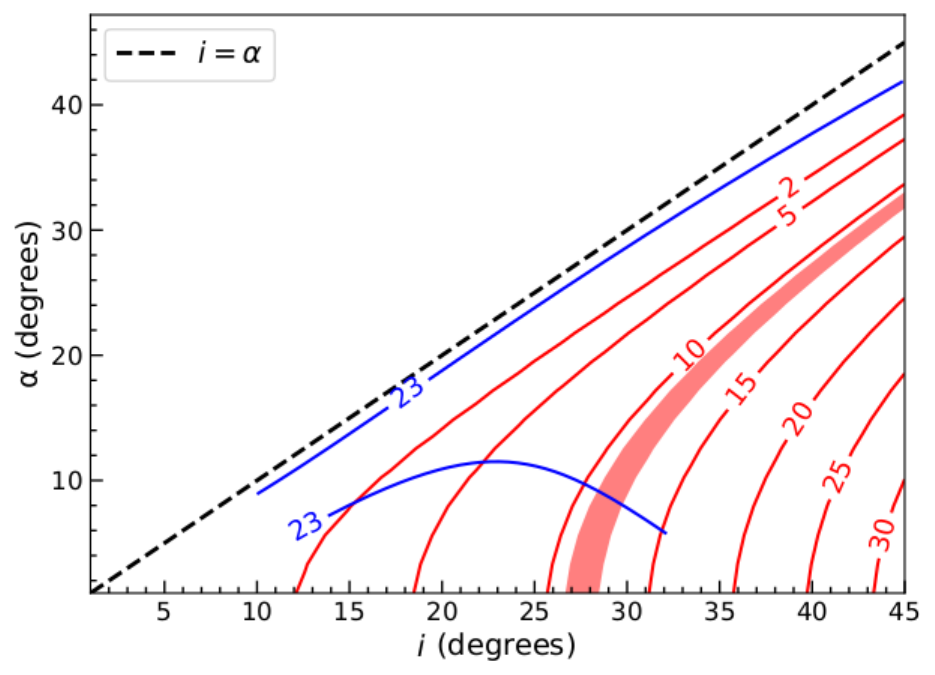}
    \caption{Contours of constant PD for radiation undergoing single scattering from the funnel volume having half opening angle $\alpha$ and different observer inclinations $i$, are depicted in red. The top panel shows results where absorption after scattering is neglected, while the bottom panel includes the effects of absorption. The red band shows the polarization degree range of 10.8$^\circ$-12.4$^\circ$ and the contour of constant PD of 23$\%$ for the case of reflection from the inner wall is shown with blue contour.}
    \label{fig:scattering_incl30}
\end{figure}


The high PD of Cyg X-3 and low apparent luminosity is attributed to significant Thomson scattering from funnel-like region surrounding the disk ({\citealt{Veledina2024}}, \citealt{veledina2024ultrasoftstatemicroquasarcygnus}). Figure \ref{fig:fig5} illustrates the funnel geometry for the scattering model, where scattering occurs within the gas present in the funnel's volume. 
\\
The detailed derivation for the observed flux and polarization for different funnel parameters are shown in the Appendix. The geometry and the results are the same as described by \cite{Veledina2024} except for the scattering case, where we also include the effect of absorption taken after the scattering. 
\section{Luminosity Estimation}

The viewing angle of the reflecting surface significantly affects the reflected spectrum from a plane disk. The scattered flux received by an observer at an inclination $i$ in a non-relativistic regime corresponding to a semi-infinite plane parallel slab irradiated by optically thin corona above the slab (\citealt{1995MNRAS.273..837M}) is :
\[
F_{s} = \frac{3 \lambda}{16}    \mu [(3- 2\mu^2 + 3\mu^4) \log(1+ \frac{1}{\mu}) + (3\mu^2 - 1)(\frac{1}{2}+\mu)]   F_{i}'
\tag{1}
\label{eq1}
\]

where, $\lambda$ is the single scattering albedo, which is the ratio of scattering efficiency to the total extinction efficiency, $\mu$ is the cosine of inclination and $F_{i}'$ is the intrinsic flux of the compact object situated above the accretion disk.\\
We define a scattered flux ratio ($R_{d}$) for plane disk model as:
\[
R_{d}= \frac{F_{s}}{\lambda F_{i}'}
\tag{2}
\label{eq:2}
\]
For $\mu$=0.866 corresponding to 30$^\circ$ observer inclination, we obtain: 
\[
R_{d}=0.323
\]
We define another scattered flux ratio ($R_{f,s}$) for scattering model of funnel as:
\[
R_{f,s}=\frac{F_{s,s}}{\lambda F_{i,s}}
\tag{3}
\label{eq:ratio}
\]
Where, $F_{i,s}$ and $F_{s,s}$ are the intrinsic flux of the source and scattered flux respectively for the scattering case.
We note that the observed scattered flux should be equal in both cases, i.e., $F_s$ = $F_{s,s}$.
We consider one example of the funnel model for the case of scattering (including absorption before and after the scattering) taking place from the gas present inside the funnel with an optical thickness corresponding to the radius of the base of the funnel ($\tau_{\rho}$) = 0.05 and $\xi$ = $13^{\circ}$ which resulted in PD of 10.7 $\%$. The scattered flux ratio for this configuration was obtained using the funnel scattering model:
\[
R_{f,s}=0.01299
\]
We define $\eta_{s}$ as the ratio of $R_{d}$ and $R_{f,s}$ representing the ratio of intrinsic fluxes of the source for plane and funnel models:  
\[
\eta_{s}= \frac{R_{d}}{R_{f,s}}
\]
We calculate $\eta_{s}$ for this configuration:
\[
\eta_{s}= \frac{0.323}{0.01299} = 24.9
\]
The intrinsic flux for the scattering model can now be written as:
\[
F_{i,s} = \eta_{s} F_{i}' \approx 25 F_{i}' 
\tag{4}
\label{eq:lum}
\]
The intrinsic flux in the case of scattering from funnel scattering model is 25 times higher than the case of a semi-infinite plane-parallel disk model. Consequently, the luminosity will also be 25 times higher.
The plane disk model used in the analysis results in an unabsorbed flux of \( 2.56 \times 10^{-7} \) erg/s/cm\(^2\). The corresponding luminosity (assuming the distance 9.7 kpc) will be:
\[
L' = F_{model} \cdot 4 \pi D^2 = 2.88 \times 10^{39} erg/s
\tag{5}
\]
and the corresponding Luminosity for the case of funnel model with scattering:
\[
L_s = 7.2 \times 10^{40} erg/s
\tag{6}
\]

Figure \ref{fig:scattering_incl30} presents the polarization degree (PD) for various observer inclinations and funnel opening angles, with contours of constant PD for the scattering model shown in red. When absorption after scattering is ignored, our results align with \citealt{veledina2024ultrasoftstatemicroquasarcygnus}, as shown in the top panel, where there is only a small change in PD with increasing funnel opening angle. In contrast, when absorption after scattering is considered, there is a significant drop in PD with increasing funnel opening angle, as shown in the bottom panel of Figure \ref{fig:scattering_incl30}. Table \ref{tab:scattering} lists the estimated intrinsic luminosity of the source for fixed observer inclination of 30$^\circ$ and different funnel parameters using scattering model while considering absorption after scattering. Additionally, Table \ref{tab:reflection} lists the estimated intrinsic luminosity of the source for different observer inclinations and funnel parameters using the reflection model.\\

\setlength{\tabcolsep}{12pt}
\begin{table*}
 \caption{Estimated luminosity of the source in case of scattering from the volume of funnel for different funnel parameters at fixed observer inclination of 30$^\circ$. (1) Optical thickness corresponding to the radius of the base of the funnel; (2) Optical thickness corresponding to the height of obstruction measured from the base; (3) Semi vertical opening angle of the funnel; (4) Observed polarization degree corresponding to the parameters of funnel ; (5) Model flux calculated using XSPEC model `cflux' in 1.0-20.0 keV; (6) Scattered flux ratio for the case of reflection from plane disk illuminated by an isotropic source present above the disk; (7) Scattered flux ratio for the case of scattering from the volume of funnel; (8) Ratio of intrinsic flux for plane disk model to the funnel scattering model (9) Intrinsic luminosity of the source for scattering model obtained by comparing the plane disk model and scattering model.}

    \centering
    \begin{tabular}{c c c c c c c c c}
    \hline \hline
    $\tau_{\rho}$ &  $\tau_{z,min}$  & $\alpha$  & PD & F$_{model}$ & $R_{d}$  & $R_{f,s}$ & $\eta_{s}$ & Luminosity  \\
     & & (Degrees) & & (10$^{-7}$$erg/s/cm^{2}$) &  & (10$^{-2}$)& & (10$^{40}$ erg/s)\\
     (1) & (2) & (3) &(4) & (5) & (6) & (7) & (8) & (9)\\
\hline 
0.01 & 0.1 & 12 & 12.4 & 2.56 & 0.323 & 0.965 & 33.5 & 9.62  \\
0.01 & 0.1 & 16 & 10.8 & 2.56 & 0.323 & 1.303 & 24.8 & 7.13\\
\hline
0.01 & 0.2 & 11 & 12.5 & 2.56 & 0.323 & 0.704 & 45.9 & 13.19  \\
0.01 & 0.2 & 15 & 10.9 & 2.56 & 0.323 & 0.987 & 32.7 & 9.41\\
\hline
0.05 & 0.2 & 10 & 12.0 & 2.56 & 0.323 & 1.066 & 30.3 & 8.71  \\
0.05 & 0.2 & 13 & 10.7 & 2.56 & 0.323 & 1.299 & 24.9 & 7.15\\

\hline
    \end{tabular}

    \label{tab:scattering}
\end{table*}

\begin{table*}

\caption{Estimated luminosity of the source in case of reflection taking place from the walls of funnel shaped geometry for different funnel parameters corresponding to 23$\%$ PD. (1) Inclination of the observer relative to the disk; (2) Model parameter R which represents the distance from the central source to the upper edge of the funnel (in units of radius of the base of funnel); (3) Semi vertical opening angle of the funnel; (4) Model flux calculated using XSPEC model cflux in 1.0-20.0 keV; (5) Scattered flux ratio for the case of reflection from plane disk illuminated by an isotropic source present above the disk; (6) Reflected flux ratio for the case of reflection from inner walls; (7) Ratio of intrinsic flux for plane disk model to the funnel reflection model ; (8) Intrinsic luminosity of the source for reflection model obtained by comparing the plane disk model and funnel reflection model.}

    \centering
    \begin{tabular}{c c c c c c c c}
    \hline \hline
    Inclination &  R  & $\alpha$  & F$_{model}$ & $R_{d}$  & $R_{f,r}$ & $\eta_{r}$ & Luminosity  \\
    (Degrees) && (Degrees) & (10$^{-7}$$erg/s/cm^{2}$) &  & (10$^{-2}$) && (10$^{40}$ erg/s)\\
    (1) & (2) & (3) & (4) &(5) & (6) & (7) & (8)\\
\hline
30 & 10 & 7.8 & 2.56 & 0.323 & 0.1769 & 182.6 & 52.5  \\
25 &10 & 11.2 & 2.53 & 0.326 & 0.7301 & 44.6 &12.7 \\
20 &10 & 10.9 & 2.53 & 0.329 & 1.4241 &23.1& 6.6 \\
\hline
30 & 20& 9.4 & 2.56 & 0.323 & 0.1496 & 215.9 & 62.1 \\
25 & 20 & 15.2 & 2.53 & 0.326 & 0.9432 & 34.6 & 9.8 \\
20 & 20 & 14.9 & 2.53 & 0.329 & 2.3575 & 13.8 & 4.0 \\
\hline
30 & 30 & 10.4 & 2.56 & 0.323 & 0.1225 & 263.6 &75.8 \\
25 & 30& 17.5 & 2.53 & 0.326 & 1.0873 & 29.9 & 8.5 \\
20 & 30& 16.5 & 2.53 & 0.329 & 2.8054 & 11.7 &3.3 \\

\hline
    \end{tabular}

    \label{tab:reflection}
\end{table*}

\section{Summary and Discussions}
In this work, we performed spectral and timing analysis of Cygnus X-3, employing simultaneous data obtained from \emph{AstroSat} and \emph{Insight}-HXMT. We divided the data of one complete orbit into three orbital phases namely; Declining phase, Minimum phase, and Rising phase. We performed spectral analysis in a broad 1.0-20.0 keV energy band. The spectral fitting analysis indicates that the source exhibits a strong dominant reflection of disk blackbody and Comptonized emission. The source was in its soft X-ray state during the observation as observed from HID and significantly low cutoff and steep power law index ($\sim$ 5.7) in spectral fitting. The spectral parameters exhibited stability throughout the orbital motion of the source, showing no significant variations except for the normalization. The normalization increases by almost a factor of 3 during orbital motion from superior to inferior conjunction. The lack of variation in the column density suggests that the orbital variation of flux can not be caused by line-of-sight absorption variations of the ionized gas (\citealt{Kallman_2019}) and asymmetric geometry of the reflector might be responsible for the modulations (\citealt{1982ApJ...257..318W}). 
\\
The choice of a purely reflection-based modeling was driven by the observed high PD. Interestingly, the observed iron line (equivalent width  $\sim$ 0.1 keV) appears significantly weaker than expected for a purely reflected spectrum, where a stronger line with an equivalent width around 1 keV is typically predicted, especially in the soft state. The observed iron line weakness may be connected to the inferred iron depletion in the stellar wind of Cygnus X-3, estimated to be around 0.1 to 0.5 times the cosmic abundance, which aligns with the system’s classification as a nitrogen-rich Wolf-Rayet (WNE) companion star (\citealt{1995A&A...294..443T}). However, the explanation for the low iron line equivalent width as being due to an underabundance of iron may not be satisfactory if the line is observed to be strong in other spectral states. \textbf{Thus, although a pure reflection spectral model is motivated by the observed large polarization degree, the weakness of the iron line may indicate that the fitting maybe of a phenomenological nature rather than a physically motivated one.} \\
 We considered a funnel geometry for two cases. In the first case, we consider scattering from the volume of the funnel while absorption after scattering is neglected, we obtained results consistent with \cite{veledina2024ultrasoftstatemicroquasarcygnus}, showing that the polarization degree (PD) does not decrease significantly with the funnel's opening angle. Interestingly, when absorption is accounted for in the second case, considering the entire volume before and after scattering, we observed a significant drop in PD with the widening of the funnel. Scattered photons from near the funnel wall have a different polarization degree and angle as compared to those which are scattered from the axis, due to the different scattering angle. Since the scattered photons from near the funnel wall, experience less absorption than those coming from the axis, the inclusion of absorption changes the net polarization degree. By calibrating our model to match the observed PD of approximately 12$\%$, we derived an intrinsic luminosity of $\sim$ 7 $\times$ 10$^{40}$ \text{ erg/s}. However, the maximum polarization produced by scattering at an observer inclination of 30$^\circ$ is 14.29$\%$ which would be produced from an infinitely narrow cylinder, indicating that the higher observed PD to the hard state cannot be explained by this scenario alone (\citealt{veledina2024ultrasoftstatemicroquasarcygnus}). \\

The modeling of the funnel in both cases indicates a super-Eddington nature for the source with L $\sim$ 10$^{40}$ erg/s. Interestingly, we can observe from Figure \ref{fig:scattering_incl30} that there is an overlap between 23$\%$ PD in the reflection model and 10.8-12.4$\%$ PD band of scattering model at inclination close to 30$^\circ$. These common funnel parameters, specifically the inclination and funnel opening angles, effectively explain the observed polarization trends in Cygnus X-3. In the scattering model, these parameters account for the lower polarization degrees typically observed in the soft state. Conversely, in the reflection model, they successfully explain the higher polarization degrees associated with the hard state. The observations used for this study are not simultaneous with the IXPE observations. Given the variability of the source, it is essential to have polarimetric measurements with broadband spectra.  

\section*{acknowledgments}
SKC acknowledges the support from UGC, Government of India for providing fellowship under the UGC-JRF scheme (NTA Ref. No.: 201610266337), and Inter-University Centre for Astronomy and Astrophysics (IUCAA), Pune for the IUCAA Visitors Programme. GM acknowledges the support from the China Scholarship Council (CSC), Grant No. 2020GXZ016647, and IUCAA Visitors Programme. AP acknowledges financial support from the IoE grant of Banaras Hindu University (R/Dev/D/IoE/Incentive/2021-22/32439), financial support through the Core Research Grant of SERB, New
Delhi (CRG/2021/000907) and thanks the Inter-University Centre for Astronomy and Astrophysics (IUCAA), Pune for associateship.
\begin{acknowledgments}

\end{acknowledgments}

%

\vspace{5mm}
\facilities{\emph{AstroSat}(LAXPC and SXT), \emph{Insight}-HXMT(LE and ME)}


\software{Heasoft \cite{heasoft_reference}, LaxpcSoft}



\bibliography{bib}{}
\bibliographystyle{aasjournal}

\begin{appendix}
\renewcommand{\thefigure}{A\arabic{figure}}
\section{Analytical modeling of funnel geometry}
\label{sec:introduction}

\label{sec:theory}
\begin{figure}[h]
    \centering
    \includegraphics[width=2.8in,angle=0,trim=0 0 0 0,clip]{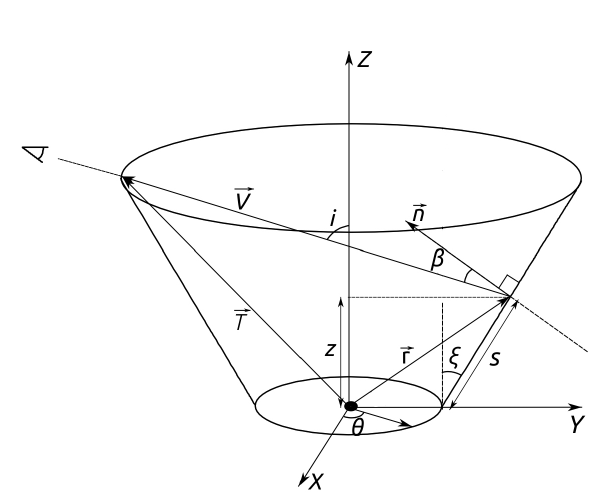} \includegraphics[width=2.8in,angle=0,trim=0 0 0 0,clip]{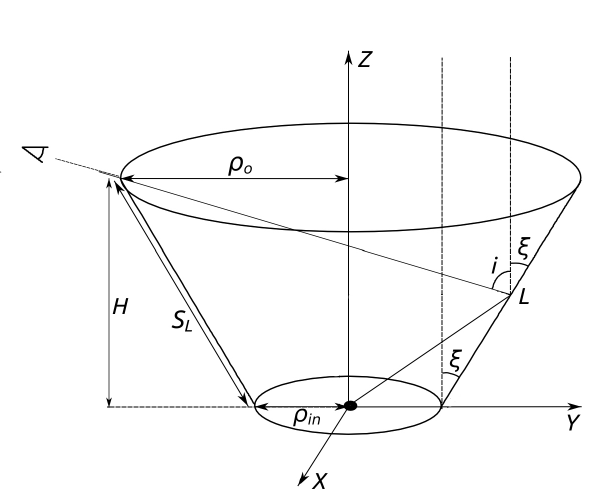}  
    \caption{Top and bottom panel displays parameters of the funnel and vectors for an arbitrary point on the funnel wall. The lowest visible point is marked as point $L$ on the funnel wall for fixed observer inclination $i$.}
    \label{fig:fig2}
\end{figure}

This work includes the funnel geometry proposed by \citealt{Veledina2024}, which suggests that a funnel with small opening angles can account for the high polarization observed in Cygnus X-3. \\
We consider a funnel-shaped structure, where the central source having radius $\Gamma$ is positioned at the bottom center. The funnel has an inner radius \(\rho_{in}\), an outer radius \(\rho_{o}\), a vertical height \(H\), and a slant height \(S_L\) as shown in bottom panel of Figure \ref{fig:fig2}. The flux emitted by the central source is reflected from the inner walls or scattered from the interior, eventually emerging through the funnel's opening. This modified flux is observed by an observer situated at a large distance \(d\) and at an inclination angle \(i\) with respect to the axis of the funnel. In this section, we calculate the flux and polarization observed by the observer and estimate the intrinsic luminosity of the source.

We first consider the reflection scenario with central source placed at the origin and the funnel opening along positive Z axis as shown in Figure \ref{fig:fig2}. 
The position vector of an arbitrary scattering point on the wall can be written as,
\[
\emph{\textbf{r}} = r\cos\theta \emph{\textbf{i}} + r\sin \theta \emph{\textbf{j}} + z \emph{\textbf{k}}
\]
Here, \( \emph{i} \), \( \emph{j} \), and \( \emph{k} \) represent the unit vectors along the \( X \), \( Y \), and \( Z \)-axes, respectively. The angle \( \theta \) is the angle subtended by the vector \( \emph{r} \) in the \( X \)-\( Y \) plane, measured counterclockwise from the positive \( X \)-axis.
\\
The $Z$ component of $\emph{\textbf{r}}$ can be expressed in terms of corresponding slant length $s$ and the funnel opening angle $\xi$ as, 
\[
z = s \cos\xi
\]
The azimuthal component of $\emph{\textbf{r}}$ is a function of $z$:
\[
r = \rho_{\text{in}} + z \tan{\xi} = \rho_{\text{in}} + s \sin{\xi}
\tag{A1}
\label{eq:6}
\]
The vector oriented in the direction of the observer at an inclination angle $i$, is given by:
\[
{\emph{\textbf{V}}} = - V_r \emph{\textbf{j}} + \frac{V_r}{\tan(i)} \emph{\textbf{k}}
\]
Let us introduce a vector \(\emph{\textbf{T}}\) such that:
\[
\emph{\textbf{T}} = \emph{\textbf{r}} + \emph{\textbf{V}} = r\cos\theta \emph{\textbf{i}} + (r\sin \theta - V_r) \emph{\textbf{j}} + (z + \frac{V_r}{\tan(i)}) \emph{\textbf{k}}
\tag{A2}
\label{eq:7}
\]
At the lowest point of visibility ($L$), the azimuthal component of vector \(\emph{\textbf{T}}\) must be equal to the outer radius of the funnel. i.e., $T_r = \rho_\text{o}$, which yields:
\[
T{_r}^{2} = (r\cos\theta)^{2} + (r\sin \theta - V_r)^{2} = \rho^2_{o}
\]
Solving above equation for $V_r$:
\[
V_{r} = r \sin\theta \pm \sqrt{ (\rho^2_{o} - r^2\cos^2\theta)}
\tag{A3}
\label{eq:8}
\] 
where we consider only the positive root as it provides the correct solution. To determine all visible points on the inner wall, we use the necessary condition that the $Z$ component of $\emph{\textbf{T}}$ must be greater than the height of the funnel in order to be visible to the observer. i.e., $T_z \geq H$:

\[
z \geq H- \frac{(V_r)}{\tan{i}}
\]

\begin{figure}
    \centering
    \includegraphics[width=3in,angle=0,trim=0 0 0 0,clip]{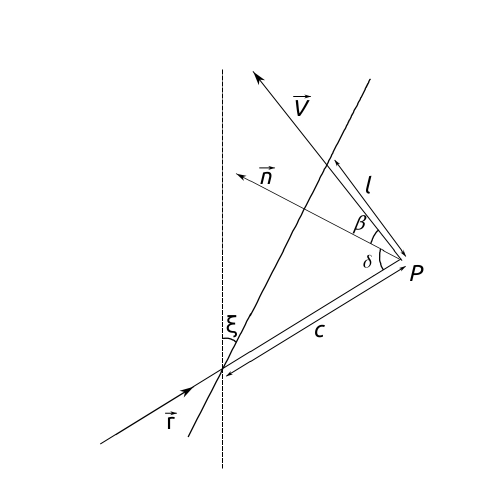}
    \caption{Above figure shows the path traced by the incident and the scattered radiation beam from the wall of the funnel. The inclined solid line depicts the funnel wall. The incident ray travels distance $c$ before scattering at point $P$ and continues for an additional distance $l$ before exiting the wall. }
    \label{fig:scattering}
\end{figure}

By substituting the values from Equations.~\eqref{eq:6}, \eqref{eq:7}, and \eqref{eq:8}, above equation for the lowest visible point transforms into a quadratic equation of the form:
\[
ps^2 + qs +r = 0
\]
where,
\[
p = \cos^2{\xi}\tan^2{i} + \sin^2{\xi} + 2\sin{\xi}\cos\xi\tan{i}\sin\theta
\]
\begin{equation}
	\begin{split}
		q = 2(\rho_{in}\sin\xi + \rho_{in}\cos\xi\tan{i}\sin\theta \\
		\quad - S_{L}\tan^2{i}\cos^2\xi - S_{L} \sin\xi\cos\xi\tan{i}\sin\theta)
	\end{split}
\end{equation}

\[
r = \rho_{in}^2 + S_{L}^2\tan^2{i}\cos^2\xi - 2S_{L}\rho_{in}\tan{i}\cos\xi\sin\theta - \rho_{o}^2
\]
The solution of the above equation provides us with the lower visible limit on the slant length of the funnel wall:
\[
s_1 = \frac{-q - \sqrt{q^2-4pr}}{2p}
\tag{A4}
\label{eq:9}
\]
The upper limit of $s$ is the upper edge of the funnel. i.e., $s_2$  = $S_{L}$.
\\
The radiation does not simply reflect from the surface of the wall; instead, it penetrates some depth depending on the optical thickness of the medium. We consider that a ray enters the wall, travels a distance $c$ before scattering occurs at point $P$, and then travels an additional distance $l$ before emerging from the wall as shown in Figure \ref{fig:scattering}. We now establish a relationship between the two lengths, $c$ and $l$.

The angle between the normal to the surface and the observer is different for different points on the wall of the funnel. 
The normal vector to any point on the inner wall of the funnel:
\[
\emph{\textbf{n}} = - r\cos\theta \emph{\textbf{i}} - r\sin\theta \emph{\textbf{j}} + {r\tan\xi} \emph{\textbf{k}}
\]

The angle $\beta$ between the vectors $\emph{\textbf{V}}$ and $\emph{\textbf{n}}$ can be computed as:

\[
\cos\beta = \frac{\emph{\textbf{n}} \cdot \emph{\textbf{V}}}{\left| \emph{\textbf{n}} \right| \left| \emph{\textbf{V}} \right|} = (\sin\theta \tan{i} + \tan{\xi})   \cos{i} \cos{\xi}
\tag{A5}
\label{eq:10}
\]

Also, angle $\delta$ between the vectors $\emph{\textbf{r}}$ and $\emph{\textbf{n}}$ :

\[
\cos\delta = \frac{\emph{\textbf{n}} \cdot \emph{\textbf{r}}}{\left| \emph{\textbf{n}} \right| \left| \emph{\textbf{r}} \right|} = \frac{-\rho_{in}}{\sqrt{r^{2}+z^{2}}   \sec{\xi}}
\]

The components of $l$ and $c$ along the normal will be equal, allowing us to establish a relationship between the two as follows:
\[
\frac{c}{l} = \frac{ \left| \cos\beta \right|}{\left| \cos\delta \right|} = \frac{(\sin\theta \sin{i} + \cos{i}\tan\xi)  \sqrt{r^2 + z^2}}{\rho_{in}}
\tag{A6}
\label{eq:11}
\]
The incident photon on the inner wall gets scattered by scattering angle $\psi$ from the initial direction.
\[
\cos\psi = \frac{\emph{\textbf{r}} \cdot \emph{\textbf{V}}}{\left| \emph{\textbf{r}} \right| \left| \emph{\textbf{V}} \right|}
\]
\[
\cos\psi = \frac{z \cos{i} - r\sin\theta \sin{i}}{\sqrt{r^2+z^2}}  
\tag {A7}
\label{eq:12}
\]

\subsection{Intensity After Scattering}
\label{sec:polarization}

We now use the radiative transfer equation to estimate the intensity of a scattered ray as it interacts with the funnel wall. As the ray incident on the wall undergoes scattering and emerges, its intensity decreases due to absorption and scattering along its path. Additionally, the intensity is augmented since each point along the path acts as a new scattering center, contributing an emission component to the ray. \\
Assuming the initial intensity of radiation from the source is $I_0$, the Flux at the point of scattering $P$ will be:
\[
F_p = I_0   \frac{\pi \Gamma^2}{(r^2 + z^2)}   e^{-\alpha c}
\tag{A8}
\label{eq:13}
\]
Now, as this beam moves along $l$, multiple scatterers contribute to the incident beam and we have one emission fraction as well,

The radiative transfer equation along $l$ starting from the point of emergence :
\[
\frac{dI_s}{dl} = \alpha I - j
\tag{A9}
\label{eq:14}
\]

where $\alpha= n(\sigma_{a}+\sigma_{s}) $ and $j$ are the extinction and emission coefficients respectively. Here, $\sigma_{s}$ and $\sigma_{a}$ are the scattering and absorption cross-sections, respectively, and $n$ is the number density of scattering particles. \\
The emission coefficient is defined as the energy transferred per unit time per unit solid angle per unit volume and can be written as:
\[
j = \frac{dE}{d\Omega dV dt} = n  F_p   \frac{d\sigma}{d\Omega}
\tag{A10}
\label{eq:15}
\]
where $\frac{d\sigma}{d\Omega}$ is the differential cross-section, representing the probability of scattering per unit solid angle.  \\
\[
\frac{d\sigma}{d\Omega} = \frac{3\sigma_s}{16\pi} (1+ \cos^2\psi)
\]
Using Equations~\eqref{eq:13} and~\eqref{eq:15}, we get:
\[
j = n  I_0   \frac{\pi \Gamma^2}{(r^2 + z^2)}   e^{-\alpha c}   \frac{3\sigma_{s}}{16 \pi}(1+\cos^2{\psi}) = B   e^{-\alpha c} (say)
\tag{A11}
\label{eq:16}
\]
Now we assume a trial solution for Equation~\eqref{eq:14}: 
\[
I = I_s   e^{- \gamma l}
\]
Substituting this solution in Equation~\eqref{eq:14}, we get:
\[
(\alpha +\gamma) I_{s} e^{-\gamma l} = B   e^{-\alpha c}
\]
A solution exists when,
\[
\alpha c = \gamma l \quad \text{and} \quad I_{s} = \frac{B}{\alpha + \gamma}
\]
We get the solution for the scattered intensity:
\[
I_{s,r}= \frac{3 \lambda}{16 \pi }  \frac{\pi \Gamma^2}{(r^2 + z^2)} \frac{I_{0}}{(1+\frac{c}{l})}(1 + \cos^2\psi)
\tag{A12}
\label{eq:17}
\]
where $\lambda$ represents the scattering albedo, defined as the fraction of the total extinction due to scattering:
\[
\lambda = \frac{\sigma_{s}}{\sigma_{a}+ \sigma_{s}}
\]

\subsection{Flux Estimation for Reflection Model}
The flux scattered from an infinitesimal area $dA$ of the wall reaching to the observer situated at distance $d$ at inclination $i$:
\[
dF_{s,r} = I_{s,r}   d\Omega_v = I_{s,r}   \frac{dA}{d^2}   \cos\beta = I_{s,r}   \frac{r d\theta ds}{d^2}   \cos\beta
\]
Substituting the value of $I_{s,r}$ from Equation~\eqref{eq:17}:
\[
dF_{s,r} = \frac{3 \lambda}{16 \pi }  \frac{F_{i,r}}{(r^2 + z^2)}\frac{(1 + \cos^2\psi)  \cos\beta}{(1+ \frac{c}{l})} {r d\theta ds} 
\]
Here, $F_{i,r}$ represents the intrinsic flux of the source that would reach the observer in the absence of the funnel geometry.

Integrating above equation over the visible surface gives the total scattered flux for given parameters of the funnel and inclination of the observer: 
\[
F_{s,r} =\int\limits_{-\theta_{o}}^{\pi+\theta_{o}}\int\limits_{s_{1}}^{s_{2}} \frac{3 \lambda}{16 \pi }  \frac{F_{i,r}}{(r^2 + z^2)}  
    \frac{(1 + \cos^2\psi)  \cos\beta}{(1+ \frac{c}{l})} {r d\theta ds}
\tag{A13}
\label{eq:18}
\]
where $\theta_{\circ}$ represents the angle at which the contour of the lowest visible point intersects the upper edge of the funnel and its value can be obtained using Equation~\eqref{eq:9}.
\subsection{Polarization Calculation for Reflection Model}
\label{sec:polarization reflection}

\begin{figure}
    \centering
    \includegraphics[width=3in,angle=0,trim=0 0 0 0,clip]{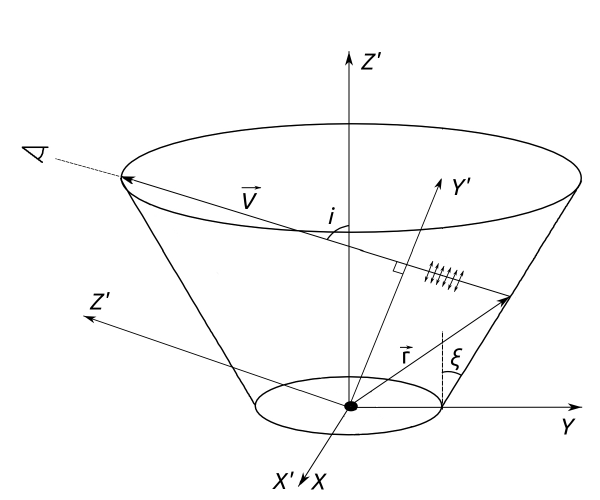}
    \includegraphics[width=3in,angle=0,trim=0 0 0 0,clip]{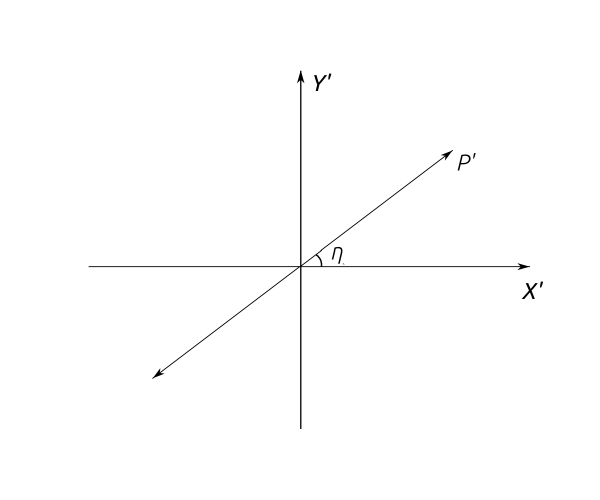}  
    \caption{The top panel shows the funnel with new coordinate after the rotation and the polarization vector of the scattered radiation and the bottom panel shows the orientation of the polarization vector in the new coordinate system.}
    \label{fig:figpolar}
\end{figure}
We introduce a vector \emph{\textbf{I}} representing the observer's position relative to the source, defined as:
\[
\emph{\textbf{I}}= -d \sin{i} \emph{\textbf{j}} + d \cos{i} \emph{\textbf{k}}
\]
where \emph{d} and \emph{i} are the distance and inclination of the observer relative to the disk. \\
The plane of polarization of scattered radiation lies perpendicular to the plane of scattering. The polarization vector lies in the direction of $\emph{\textbf{r}} \times \emph{\textbf{I}}$ and can be written as:
\begin{equation}
	\begin{split}
		\emph{\textbf{P}} = (r.d.\sin\theta\cos{i} + z.d. \sin{i})\emph{\textbf{i}} 
		- (r.d. \cos\theta\cos{i}) \emph{\textbf{j}} \\
		- (r.d. \cos\theta \sin{i}) \emph{\textbf{k}}
	\end{split}
	\tag{A14}
	\label{19}
\end{equation}

For mathematical simplification, we further rotate the $Y-Z$ plane by angle $i$ in an anti-clockwise direction using a rotation matrix to align the polarization vector in $X'$ and $Y'$ axes and position the observer in $Z'$ axis as shown in the top panel of Figure \ref{fig:figpolar}. 
The rotation matrix is given by:
\[
T =
\begin{bmatrix}
    1 & 0 & 0 \\
    0 & \cos{i} & -\sin{i} \\
    0 & \sin{i} & \cos{i}
\end{bmatrix}
\]
The polarization vector in the new coordinate system will be:
\[
\emph{\textbf{P'}} = \emph{\textbf{P}} \times T = (r.d.\cos{i}\sin\theta + z. d. \sin{i}) \emph{\textbf{i'}} - (r.d.\cos\theta) \emph{\textbf{j'}}
\tag{A15}
\label{eq:20}
\]
The scattered radiation is partially polarized and can be written as the sum of polarized and unpolarized components:
\[
I_{ST} = I_{sp} + I_{sup}
\]
The polarization degree of Thomson scattering depends on the scattering angle $\psi$ and is given by:
\[
PD = \frac{1-\cos^2\psi}{1-\cos^2\psi} 
\tag{A16}
\label{eq:21}
\]
Now, the intensity of scattered radiation can be rewritten as:
\[
I_{ST} = I_s \frac{(1 - \cos^2\psi)}{(1 + \cos^2\psi)} + 2I_s \frac{\cos^2\psi}{(1+\cos^2\psi)}
\tag{A17}
\label{eq:22}
\]

\begin{figure}
    \centering
    \includegraphics[width=3.4in,angle=0,trim=0 0 0 0,clip]{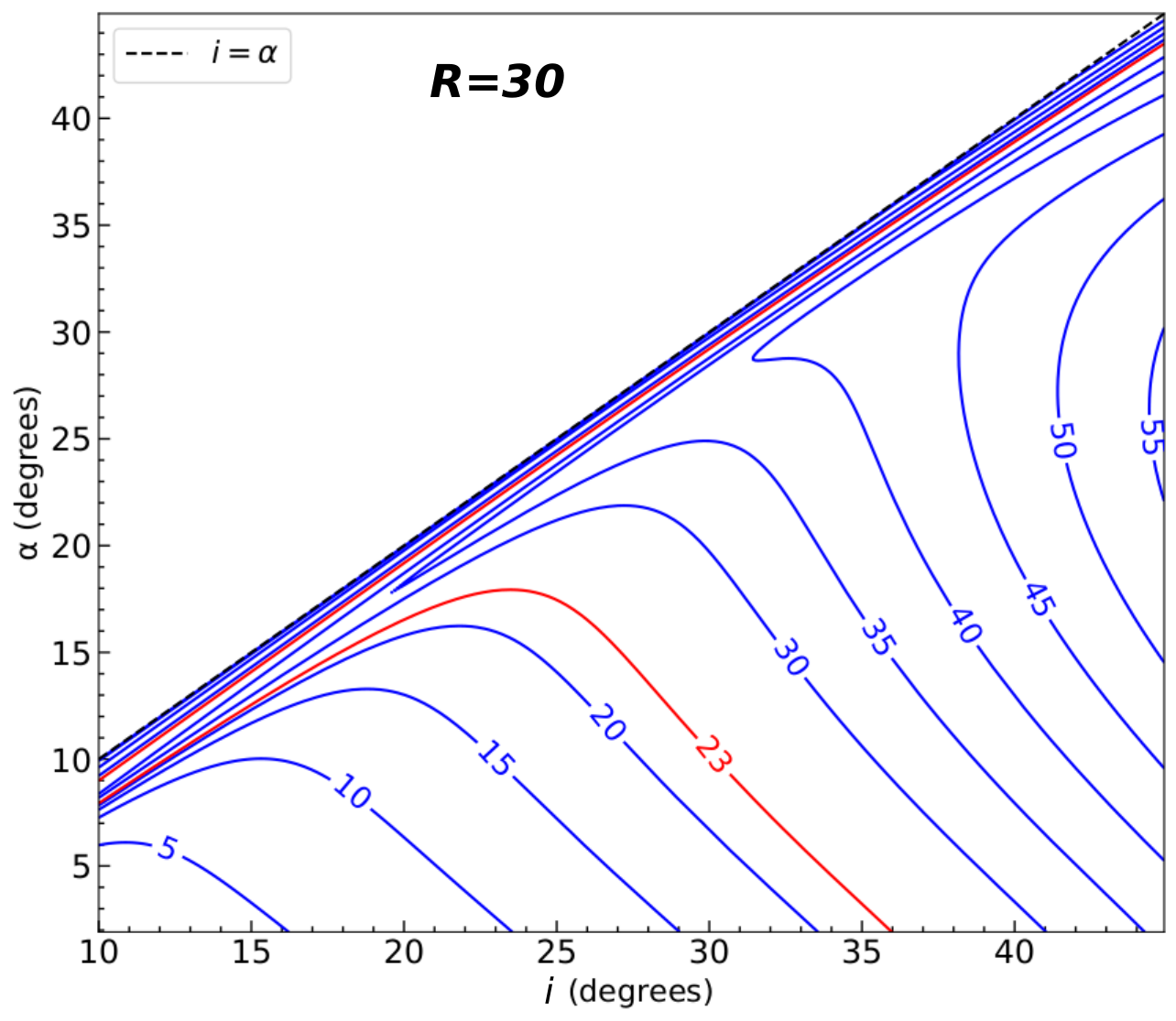} 
    \caption{Contour plot of constant PD for different observer inclinations $i$ and $\alpha$ with fixed value of $R$ for the reflection model. The definitions of R and $\alpha$ are the same as in Figure \ref{fig:fig5}.}
    \label{fig:fixed_rr_reflection}
\end{figure}

Using Malus's law, the intensity of scattered polarized radiation along $X'$ will be:
\[
I_{x'p} = I_{s}\frac{(1 - \cos^2\psi)}{(1 + \cos^2\psi)}   \cos^2\eta
\]
where $\eta$ is the angle between the polarization vector and the $X'$ axis as shown in bottom panel of Figure \ref{fig:figpolar}. Using Equation~\eqref{eq:20}, 
\[
\cos\eta = \frac{(r \cos{i} \sin\theta + z \sin{i})}{\sqrt{(r \cos{i} \sin\theta + z \sin{i})^2 + r^2 \cos^2\theta}}   
\tag{A18}
\label{23}
\]
The intensity of polarized radiation along $X'$:
\[
F_{x'p} =\int\limits_{-\theta_{o}}^{\pi+\theta_{o}}\int\limits_{s_{1}}^{s_{2}} {I_{x'p}} \, d\theta \, ds=\int\limits_{-\theta_{o}}^{\pi+\theta_{o}}\int\limits_{s_{1}}^{s_{2}} {I_{s}   \frac{1-\cos^2\psi}{1+\cos^2\psi}  \cos^2\eta}\, d\theta \, ds
\]
Similarly, the Intensity of polarized radiation along $Y'$:
\[
F_{y'p} =\int\limits_{-\theta_{o}}^{\pi+\theta_{o}}\int\limits_{s_{1}}^{s_{2}} {I_{x'p}} \, d\theta \, ds=\int\limits_{-\theta_{o}}^{\pi+\theta_{o}}\int\limits_{s_{1}}^{s_{2}} {I_{s}   \frac{1-\cos^2\psi}{1+\cos^2\psi}  \sin^2\eta}\, d\theta \, ds
\]
The PD of the observed flux will be:
\[
PD = \frac{I_{max}-I_{min}}{I_{max}+I_{min}} = \frac{I_{x'p}-I_{y'p}}{I_{x'p}+I_{y'p}+I_{sup}} = \frac{F_{x'p}- F_{y'p}}{F_{s}}
\tag{A19}
\label{eq:24}
\]
We obtain the PD for different parameters of the funnel using the above equation and the contours of constant PD for different observer inclinations ($i$) and funnel opening angle ($\alpha$) and fixed values of $R$ are shown in Figure \ref{fig:fixed_rr_reflection}.

\subsection{Flux Estimation for Scattering Model}
\label{sec:flux scattering}

Initially, we considered scattering to occur solely from the walls of the cylinder. However, changes in polarization properties suggest that scattering might also be occurring within the volume inside the funnel, in addition to the walls, leading to a reduction in the net polarization.
We now consider the second scenario where the scattering takes place from the gas present in the funnel. The funnel structure is shown in Figure \ref{fig:fig5}. The main parameters for the funnel for the scattering model are $\tau_{\rho}$, the optical thickness at the base radius of the funnel, height boundaries $\tau_{z,min}$ and $\tau_{z,max}$ representing the lower and upper limits of the funnel volume and funnel opening angle $\xi$. \\
The position vector of any arbitrary scattering point inside the funnel:
\[
\emph{\textbf{c}} = r \cos{\theta} \emph{\textbf{i}} + r \sin{\theta} \emph{\textbf{j}} + z \emph{\textbf{k}}
\]
The path length covered by the ray originating from the source before scattering, denoted as $c$, is given by:
\[
\left| c \right| = \sqrt{(r^2 + z^2)}
\]
The vector directed along the observer :
\[
\emph{\emph{\textbf{l}}} = -l_r \emph{\textbf{j}} + (\frac{l_r}{\tan{i}}) \emph{\textbf{k}}
\]
Again, we introduce $\emph{\textbf{T}}$ such that :
\[
\emph{\textbf{T}} = \emph{\textbf{l}} +\emph{\textbf{c}} = r \cos{\theta} \emph{\textbf{i}} + (r \sin{\theta} - l_r) \emph{\textbf{j}} + (\frac{l_r}{\tan{i}} + z) \emph{\textbf{k}}
\tag{A20}
\label{eq:25}
\]
The azimuthal component of $\emph{\textbf{T}}$:
\[
T_r = \rho_{in} + (z + \frac{l_r}{\tan{i}})  \tan\xi
\tag{A21}
\label{eq:26}
\]
Using Equations~\eqref{eq:25} and~\eqref{eq:26}, we obtain the path covered by each ray in the funnel medium after scattering ($l$) for different points in the funnel:
\[
p_1   l_r^2 + p_2   l_r +p_3 = 0
\]
where,
\[
l_{r} = l \sin{i}
\]
\[
p_1 = \tan^2{i} - {\tan^2{\xi}} 
\]
\[
p_2 = -2\tan{i}(r\sin\theta\tan{i} + z {\tan^2{\xi}} + \rho_{in} {\tan\xi})
\]
\[
p_{3} = (r^2 - \rho_{in}^2 - z^2 \tan\xi - 2z\rho_{in}\tan\xi)  \tan^2{i}
\]
From Equation~\eqref{eq:14}, we have the scattered intensity:
\[
dI = \alpha I    dl - j   dl
\]
Assuming the funnel to be truncated at some finite distance, the ray travels total distance $l+c$ in the volume of funnel before coming out.
We can write the intensity relation as:
\[
dI_{s,s} = n   F_p   \frac{d\sigma}{d\Omega}   e^{-\alpha{l}}   e^{-\alpha{c}}   dl
\tag{A22}
\label{eq:27}
\]
Substituting $\frac{d\sigma}{d\Omega}$ in above equation, we get:
\[
dI_{s,s} = n   \frac{I_o \pi \Gamma^2}{c^2}   \frac{3\sigma_s}{16\pi} (1+ \cos^2\psi)   e^{-\alpha{(l+c)}}   dl
\]
The scattered flux from infinitesimal volume $dV$ will be: 
\[
dF_{s,s} = {dI_s   d\Omega_v} 
\]
\[
dF_{s,s} = n   \frac{I_o \pi \Gamma^2}{c^2}   \frac{3\sigma_s}{16\pi} (1+ \cos^2\psi)   e^{-\alpha{(l+c)}}      \frac{dA_{\perp}}{D^2} dl
\tag{A23}
\label{eq:28}
\]
Where $dV = dl dA_{\perp}$ and $dA_{\perp}$ is the infinitesimal area perpendicular to the line of sight of the observer. \\  
We further transform the lengths to the optical thickness by using the following transformations:
\[
\tau_{r} = \alpha r \quad and \quad \tau_{z} = \alpha z
\]
Equation~\eqref{eq:28} transforms to:
\[
dF_{s,s} = \frac{3\lambda}{16\pi}  F_{i,s}  (1+ \cos^2\psi)   \frac{e^{-(\tau_{l}+\tau_{c})}}{(\tau_{r}^2 + \tau_{z}^2 )}    dV
\]
\[
dF_{s,s} = \frac{3\lambda}{16\pi}  F_{i,s}  (1+ \cos^2\psi)   \frac{e^{-(\tau_{l}+\tau_{c})}}{(\tau_{r}^2 + \tau_{z}^2 )}    \tau_{r}    d\tau_{r} d \theta  d\tau_{z}
\tag{A24}
\label{eq:29}
\]
Here, $F_{i,s}$ represents the intrinsic flux of the source that would reach the observer in the absence of the funnel geometry. \\
By integrating Equation~\eqref{eq:29} over the entire visible volume of the funnel, and assuming the funnel is obstructed up to a certain height from the base by imposing a lower limit on $\tau_z$, while setting the upper limit at a large value to represent an infinite height (${\tau_{z_{\text{max}}}}$= 10), we obtain the total scattered flux in the direction of the observer:
\begin{equation}
	\begin{split}
		F_{s,s} = \int_{0}^{\tau_{r_{max}}} \int_{0}^{2\pi} \int_{\tau_{z_{\text{min}}}}^{\tau_{z_{\text{max}}}} 
		\frac{3\lambda}{16\pi} F_{i,s} (1 + \cos^2 \psi) \\
		\times \frac{e^{-(\tau_l + \tau_c)}}{(\tau_r^2 + \tau_z^2)} \, \tau_r \, d\tau_r \, d\theta \, d\tau_z
	\end{split}
	\tag{A25}
	\label{eq:30}
\end{equation}

where, $\tau_{r_{max}}$ = $\tau_{\rho}$ + $\tau_{z} \tan\xi$ and
$\tau_{\rho}$ is the optical thickness corresponding to the radius of the base of the funnel.

\subsection{Polarization for Scattering Model}
\begin{figure*}
    \centering
    \includegraphics[width=7.0in,angle=0,trim=0 0 0 0,clip]{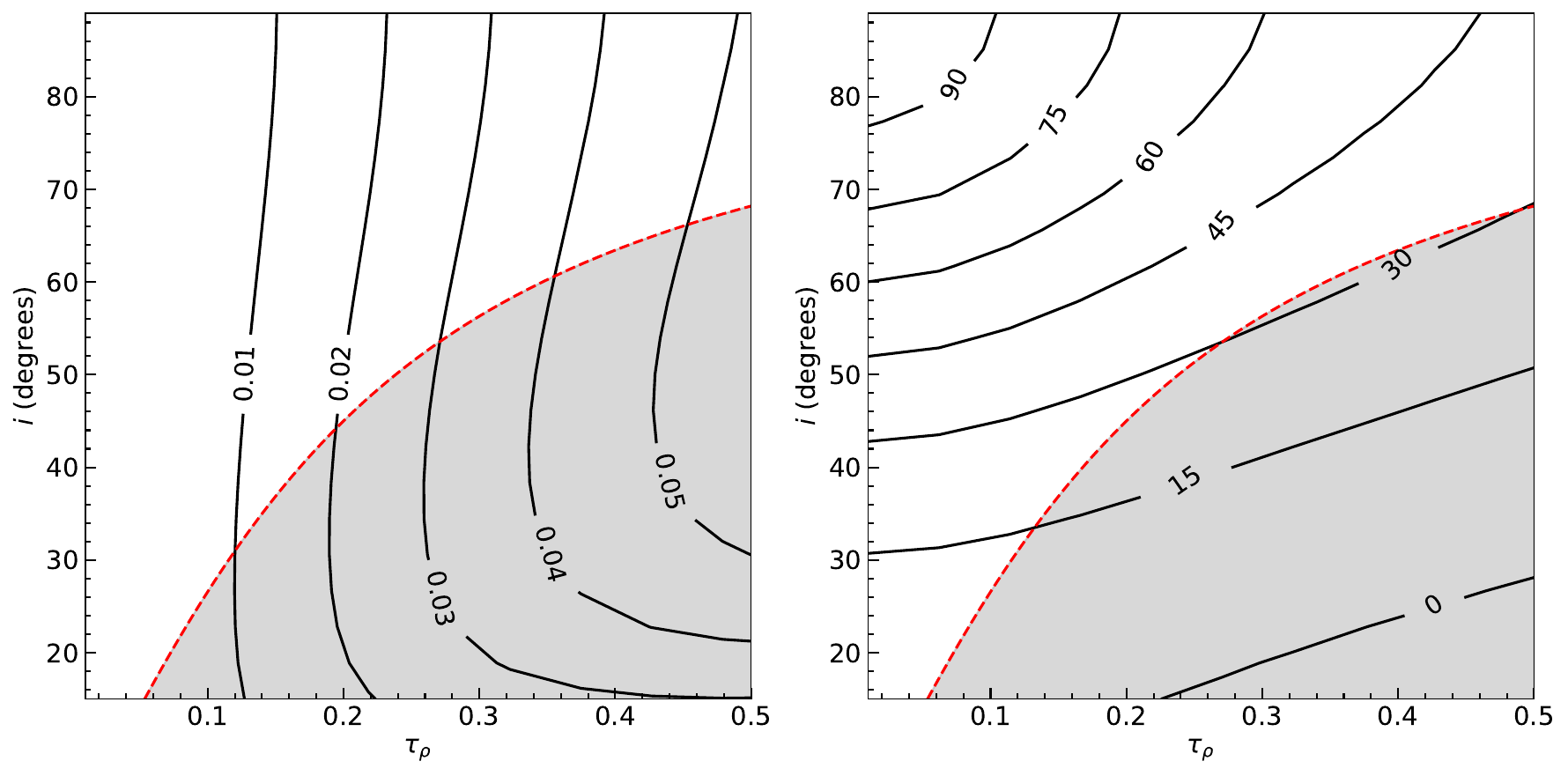}
    \includegraphics[width=7.0in,angle=0,trim=0 0 0 0,clip]{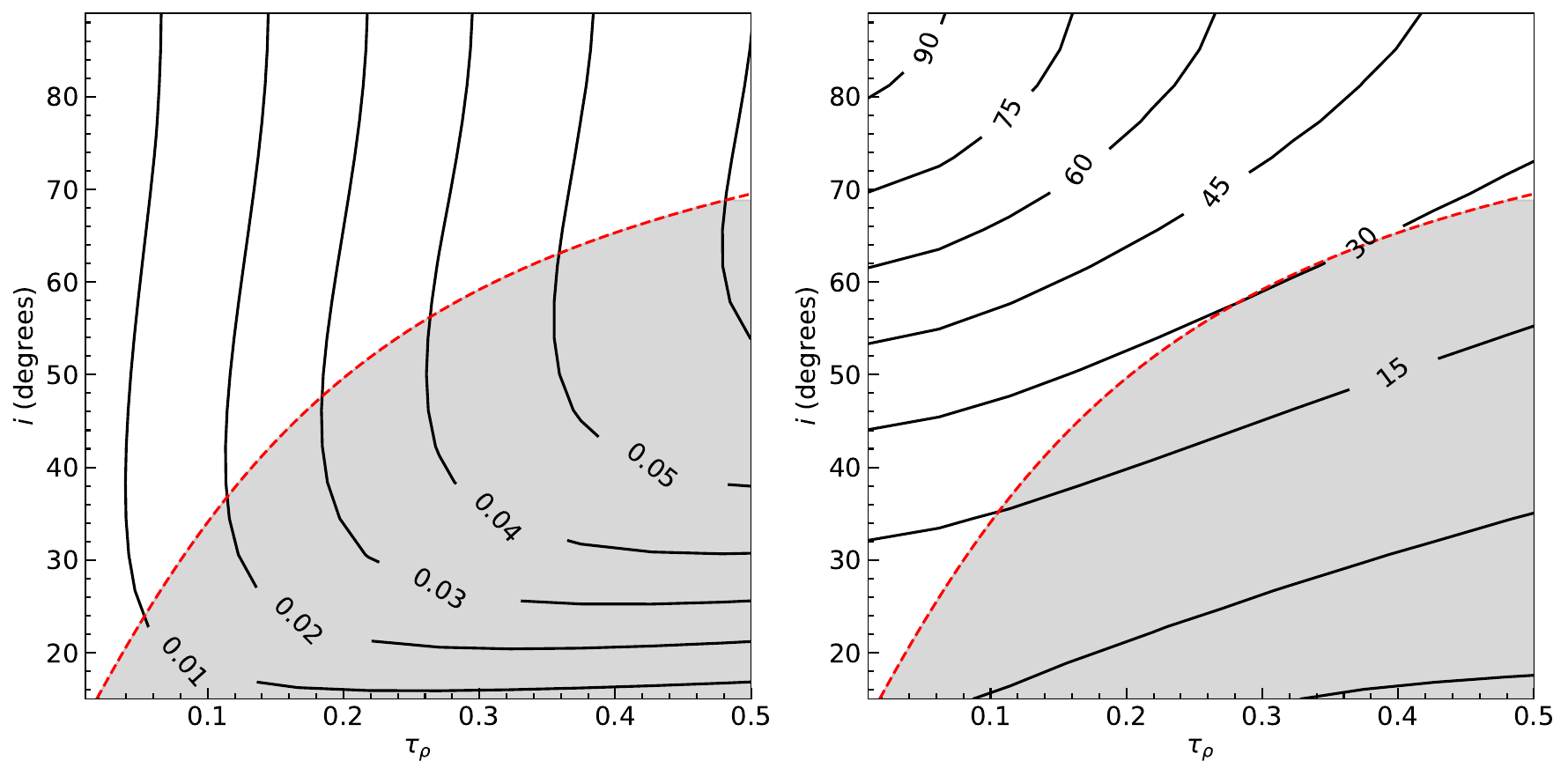}
    \caption{Left and right panels show contour plots of the constant scattered flux ratios ($R_{f,s}$) and the constant PD respectively in case of scattering taking place from the gas present in volume for various optical thicknesses and observer inclinations, with fixed funnel opening angle ($\xi$) and height boundaries ($\tau_{z,min}$ = 0.2 and $\tau_{z,max}$ = 10). The red dotted curve corresponds to the limit where the central source is visible to the observer. all the points in the shaded region correspond to the case when the central source is visible to the observer.}
    \label{fig:pd_scattering}
\end{figure*}

We use the same method as shown in section \ref{sec:polarization reflection} to calculate polarization for the scattering model. \\
The flux of polarized scattered radiation along the $X'$ axis:
\[
F_{x'p} = \int\int\int{\cos^2\eta   dF_{s,s}}
\]
Similarly, the flux of polarized scattered radiation along the $Y'$ axis:
\[
F_{y'p} = \int\int\int{\sin^2\eta   dF_{s,s}}
\]
Further, we use Equation~\eqref{eq:24} to obtain the polarization degree. The plot for contours of constant PD at different inclinations ($i$) and funnel angle ($\alpha$) for fixed $\tau_{\rho}$ (0.01) and height boundaries ($\tau_{z,min}$ = 0.2 and $\tau_{z,max}$ = 10) is shown in Figure \ref{fig:scattering_incl30}.
The contour plots of the constant scattered flux ratios (R$_{f,s}$) and the constant PD for the case of scattering taking place from the gas present in volume of the funnel for different observer inclinations and optical thicknesses at fixed $\xi$ and height boundaries ($\tau_{z,min}$ = 0.2 and $\tau_{z,max}$ = 10) are shown in Figure \ref{fig:pd_scattering}.

\subsection{Scattering Model Without Absorption}
When absorption after scattering is neglected in the scattering model, Equation~\eqref{eq:27} reduces to:
\[
dI_{s,s}' = n   F_p   \frac{d\sigma}{d\Omega}   e^{-\alpha{c}}   dl
\tag{A26}
\label{eq:31}
\]
Following a similar procedure as outlined in the previous section, we calculate the flux and polarization. The resulting contour plot for the scattering model is presented in Figure \ref{fig:scattering_incl30}.

\end{appendix}



\end{document}